# Input-State-Parameter-Noise Identification and Virtual Sensing in Dynamical Systems: A Bayesian Expectation-Maximization (BEM) Perspective

Daniz Teymouri[1], Omid Sedehi[2*], Lambros S. Katafygiotis[3], Costas Papadimitriou[4]

**Abstract**
Structural identification and damage detection can be generalized as the simultaneous estimation of input forces, physical parameters, and dynamical states. Although Kalman-type filters are efficient tools to address this problem, the calibration of noise covariance matrices is cumbersome. For instance, calibration of input noise covariance matrix in augmented or dual Kalman filters is a critical task since a slight variation in its value can adversely affect estimations. The present study develops a Bayesian Expectation-Maximization (BEM) methodology for the uncertainty quantification and propagation in coupled input-state-parameter-noise identification problems. It also proposes the incorporation of input dummy observations for stabilizing low-frequency components of the latent states and mitigating potential drifts. In this respect, the covariance matrix of the dummy observations is also calibrated based on the measured data. Additionally, an explicit formulation is provided to study the theoretical observability of the Bayesian estimators, which helps characterize the minimum sensor requirements. Ultimately, the BEM is tested and verified through numerical and experimental examples, wherein sensor configurations, multiple input forces, and abrupt stiffness changes are investigated. It is confirmed that the BEM provides accurate estimations of states, input, and parameters while characterizing the degree of belief in these estimations based on the posterior uncertainties driven by applying a Bayesian perspective.

## 1. Introduction

Vibration-based monitoring is an emerging area of research, which steps toward automated ways of inspection and monitoring in structures. The core idea is to identify dynamical properties and characterize structural damage using vibration data collected from a sensor network positioned at critical positions. While early research focused mainly on identifying abrupt changes in mechanical properties, recent works seek to track strain and stress responses in the entire body of structures [1]. This new approach goes beyond detecting drops in mechanical properties and allows capturing plastic deformations, hysteresis cycles, and fatigue damage accumulation.

Bayesian filters contribute to both perspectives of vibration-based monitoring, which activates capturing stiffness reductions and virtually sensing unmeasured responses [2–5]. The first wave of techniques, rooted in the Kalman Filter (KF), has relied on the knowledge of input forces; however, the input forces are feasible to measure only in simulations and laboratory investigations. Consequently, a new wave of filters has formed in recent years that identifies input forces from response-only measurements [6–14]. This achievement has created new opportunities for structural identification through the simultaneous estimation of mechanical properties, input forces, and unmeasured dynamical responses. This process, referred to as coupled input-state-parameter estimation, is the subject of this paper.

A general approach to the coupled estimation problems describes the temporal variation of unknown parameters and input forces through random walk models to construct a latent state vector

[1] Ph.D. Student, Department of Civil and Environmental Engineering, The Hong Kong University of Science and Technology, Hong Kong, China, **Email:** dtyemouri@connect.ust.hk

[2*] Postdoctoral Fellow, Department of Civil and Environmental Engineering, The Hong Kong University of Science and Technology, Hong Kong, China, **Email:** osedehi@connect.ust.hk **(Corresponding Author)**

[3] Professor, Department of Civil and Environmental Engineering, The Hong Kong University of Science and Technology, Hong Kong, China, **Email:** katafygiotis.lambros@gmail.com

[4] Professor, Department of Mechanical Engineering, University of Thessaly, Volos, Greece, **Email:** costasp@uth.gr

governed by a joint state-space model. Then, the Extended Kalman Filter (EKF) can estimate the unknown quantities in a real-time manner [8,15,16]. Maes et al. have shown that the association of the EKF with smoothing techniques can provide superior stability and accuracy [17]. The Unscented Kalman Filter (UKF) can be a more accurate albeit computationally costly alternative to the EKF, which outperforms the EKF in tackling severe systematical or mathematical nonlinearities [18–20]. However, in these methods, the fusion of displacement and acceleration measurements has appeared as a requisite for avoiding low-frequency drifts [8,15,16]. Dertimanis et al. have relaxed this condition by estimating the input forces through a KF, parallelized with the UKF for state-parameter estimation [21]. This strategy has shown to be a substitute for the full augmentation of the state-space models, which can mitigate low-frequency drifts when fine-tuned through L-curve methods [21]. Applications of these methods for identifying seismic excitations and updating nonlinear mechanics-based models have been the subject of recent works by Astroza et al. [22,23].

The parameterization of the input forces through time-varying autoregressive models, Gaussian process models, and colored stochastic processes can help enhance the stability of input estimations [24–26]. Moreover, adaptive calibration of the measurement noise covariance matrix allows obtaining more accurate estimates [27]. Recently, sparsity-promoting variants of the EKF have also shown success in reducing spatial-temporal variability of structural parameters [28].

Although Kalman-type filters attest to their computational efficiency, their performance highly relies on the proper choice of noise characteristics. In most practical cases, the noise parameters are scarcely known and should be calibrated based on the measurements. For this purpose, pragmatic rules of thumb enable fair-tuning of the noise parameters, e.g., [29,30]. Although these methods have been successful to some extent, the users' judgment has a dominant influence on the outcome. Optimization-based calibration of noise parameters is a more legitimate strategy, which can be performed on two grounds:
- Minimum Mean Squared Error (MMSE) estimation: covariance matching [27,31–33] and correlation-based techniques [11,34–37].
- Probability-based: Maximum Likelihood (ML) [38] and Bayesian methods [39,40].

This study focuses on probability-based Bayesian approaches, so we will further review and discuss them. The brute force maximization of the posterior distributions has recently been adopted to identify the noise parameters of the EKF through the finite difference approach [26,40,41]. Another way to update the noise parameters is to assign conjugate prior distributions to the noise covariance matrices and obtain explicit formulations for updating the noise parameters [12,28,42]. Expectation-Maximization (EM) and Variational Bayes (VB) are other alternative methods for noise identification, which approximate the logarithm of the posterior distribution with a surrogate function, whose optimum in limit coincides with those of the original distribution [43,44]. Although the latter category of methods can effectively identify the noise covariance matrices, their applications to the coupled estimation problems have not been explored adequately.

This paper expands upon the Bayesian Expectation-Maximization (BEM) methodology [45], proposed recently by the authors for the joint input-state estimation. Unlike [45], which deals with known dynamical systems, the present study tackles the simultaneous identification of input forces, dynamical states, mechanical parameters, and noise characteristics. In this respect, the present paper is a further generalization of the BEM proposed in [45]. More specifically, the proposed BEM embodies the augmented EKF and fixed-point smoother for estimating unknown quantities in real-time and updates the noise covariance matrices through explicit formulations. This process continues until convergence criteria of the EM algorithm are satisfied, ultimately providing the joint posterior distributions of latent states and noise parameters. Moreover, a steady-state algorithm is prescribed for the initial estimation of the input and state covariance matrices, aiming to enhance the convergence and stability of the main BEM algorithm. In the sense that the noise characteristics are identified from the data, the BEM well suits uncertainty quantification and propagation in coupled estimation problems. In this work, input dummy observations are included for stabilizing low-frequency components of the latent states and mitigating fictitious drifts when only acceleration responses are available. A closed-form expression is also provided for testing theoretical observability of partially known systems with unknown input using Lie derivatives. Ultimately, the BEM is applied to numerical and experimental examples, showcasing its efficacy.

Compared to references [25,46] that use Gaussian process models associated with brute-force optimization or sampling techniques, the present study employs the random walk model in conjunction with the EM algorithm. Unlike these references, the proposed methodology does not require any optimization or sampling techniques and acts as a standalone algorithm with explicit steps. Therefore, in this sense, it offers some computational savings as well.

The organization of this paper is as follows. Section 2 builds up the process and observation models and analyzes theoretical observability. The proposed Bayesian formulation is introduced in Section 3. The computational aspects appear in Section 4, and the computational algorithm is described in Section 5. The applicability and efficiency of the BEM are verified in Sections 6 and 7 using both numerical and experimental examples. Finally, conclusions and future work are explained in Section 8.

## 2. Probabilistic Process and Observation Models
### 2.1. State-space Process Model

The vibration response of a linear structure is governed by the Newton's second law of motion, which gives:

$$\mathbf{M}\ddot{\mathbf{x}}(t) + \mathbf{C}(\boldsymbol{\theta}(t))\dot{\mathbf{x}}(t) + \mathbf{K}(\boldsymbol{\theta}(t))\mathbf{x}(t) = \mathbf{S}_p \mathbf{p}(t) \tag{1}$$

where system dynamics are represented through $\mathbf{M} \in \mathbb{R}^{N_d \times N_d}$, $\mathbf{C}(\boldsymbol{\theta}(t)) \in \mathbb{R}^{N_d \times N_d}$, and $\mathbf{K}(\boldsymbol{\theta}(t)) \in \mathbb{R}^{N_d \times N_d}$, which denote the mass, damping, and stiffness matrices, respectively; $N_d$ is the total number of degrees-of-freedom (DOFs); $\mathbf{x}(t) \in \mathbb{R}^{N_d}$ is the displacement response; $\mathbf{p}(t) \in \mathbb{R}^{N_p}$ is a vector of unknown external forces or excitations; $\mathbf{S}_p \in \mathbb{R}^{N_d \times N_p}$ is a known matrix indicating the spatial distribution of $N_p$ unknown dynamical forces; the over dot indicates the derivative with respect to continuous time.

In practice, structures' mass might not undergo significant changes over time and can be estimated using engineering drawings. Thus, the mass matrix is often assumed to be fixed and known. In contrast, the stiffness and damping matrices can be partially or entirely unknown, and their unknown elements can be collected into $\boldsymbol{\theta}(t) \in \mathbb{R}^{N_\theta}$, comprising $N_\theta$ unknown structural parameters. The parameterization of the stiffness and damping matrices is important as it influences the efficiency and accuracy of the system identification. In this regard, a convenient approach is to express the total stiffness/damping matrices as a linear function of the substructural stiffness/damping matrices multiplied by unknown parameters, giving:

$$\mathbf{K}(\boldsymbol{\theta}(t)) = \mathbf{K}_0 + \sum_{s=1}^{N_k} \theta_s(t)\mathbf{K}_s \tag{2}$$

$$\mathbf{C}(\boldsymbol{\theta}(t)) = \mathbf{C}_0 + \sum_{s=N_k+1}^{N_\theta} \theta_s(t)\mathbf{C}_s \tag{3}$$

where $\mathbf{K}_0 \in \mathbb{R}^{N_d \times N_d}$ and $\mathbf{C}_0 \in \mathbb{R}^{N_d \times N_d}$ are the known parts of the stiffness and damping matrices, respectively; $\mathbf{K}_s \in \mathbb{R}^{N_d \times N_d}$ and $\mathbf{C}_s \in \mathbb{R}^{N_d \times N_d}$ are the initial stiffness and damping matrices corresponding to the $s^{\text{th}}$ substructure, respectively; the dependence on time ($t$) indicates the time-varying nature of the unknown parameters. This linear-in-parameter relationship can represent a large class of linear structures since the stiffness matrix of FE models can often be written as a linear function of the constitutive substructures [5]. However, in this paper, the primary use of this assumption is derivation of observability conditions, and it does not impose any limitations on the BEM methodology proposed hereafter. When this linearity in the parameters does not hold, the BEM remains general and applicable. However, in such cases, one needs to supply the gradients of the state-space model for a potentially nonlinear parameterization.

Given this physical setting, dynamical responses can be reformulated in the state-space form, yielding:

$$\dot{\mathbf{z}}(t) = \mathbf{A}_c(\mathbf{\theta}(t))\mathbf{z}(t) + \mathbf{B}_c\mathbf{p}(t) \tag{4}$$

where $\mathbf{z}(t) = [(\mathbf{x}(t))^T \; (\dot{\mathbf{x}}(t))^T]^T$ is the state vector; $\mathbf{A}_c(\mathbf{\theta}(t)) \in \mathbb{R}^{2N_d \times 2N_d}$ and $\mathbf{B}_c \in \mathbb{R}^{2N_d \times N_p}$ respectively denote the system and the input-to-state matrices, given by

$$\mathbf{A}_c(\mathbf{\theta}(t)) = \begin{bmatrix} \mathbf{0}_{N_d \times N_d} & \mathbf{I}_{N_d \times N_d} \\ -\mathbf{M}^{-1}\mathbf{K}(\mathbf{\theta}(t)) & -\mathbf{M}^{-1}\mathbf{C}(\mathbf{\theta}(t)) \end{bmatrix} \quad ; \quad \mathbf{B}_c = \begin{bmatrix} \mathbf{0}_{N_d \times N_p} \\ \mathbf{M}^{-1}\mathbf{S}_p \end{bmatrix} \tag{5}$$

where the subscript $c$ represents continuous-time quantities. A discrete-time representation can be obtained by considering $\Delta t$ sampling intervals and characterizing the variation of input forces through the zero-order hold assumption, which provide:

$$\mathbf{z}_k = \mathbf{A}(\mathbf{\theta}_{k-1})\mathbf{z}_{k-1} + \mathbf{B}(\mathbf{\theta}_{k-1})\mathbf{p}_{k-1} + \mathbf{v}_k^z \tag{6}$$

Here, the subscript $k$ denotes the time instant $t_k = k\Delta t$, $\forall k \in \{1,...,n\}$; $\mathbf{z}_{k-1} = \mathbf{z}(t_{k-1})$, $\mathbf{p}_{k-1} = \mathbf{p}(t_{k-1})$, and $\mathbf{\theta}_{k-1} = \mathbf{\theta}(t_{k-1})$ denote the state, input, and parameter vectors at $t_{k-1}$, respectively; $\mathbf{A}(\mathbf{\theta}_{k-1}) = e^{\mathbf{A}_c(\mathbf{\theta}_{k-1})\Delta t}$ and $\mathbf{B}(\mathbf{\theta}_{k-1}) = (\mathbf{A}(\mathbf{\theta}_{k-1}) - \mathbf{I}_{2N_d \times 2N_d})\mathbf{A}_c^{-1}(\mathbf{\theta}_{k-1})\mathbf{B}_c$ are the system and input-to-state matrices of the discrete-time representation, respectively; $\mathbf{v}_k^z \sim N(\mathbf{0}, \mathbf{Q}^z)$ is the process noise described by a zero-mean Gaussian white noise (GWN) process with $\mathbf{Q}^z \in \mathbb{R}^{2N_d \times 2N_d}$ covariance matrix. The covariance matrix of this GWN process should be identified from the data.

In this paper, the structural parameters ($\mathbf{\theta}_k$) and the unknown input loading ($\mathbf{p}_k$) are treated as time-varying processes, described using first-order random walk models. Thus, we have:

$$\mathbf{\theta}_k = \mathbf{\theta}_{k-1} + \mathbf{v}_k^\theta \tag{7}$$

$$\mathbf{p}_k = \mathbf{p}_{k-1} + \mathbf{v}_k^p \tag{8}$$

where $\mathbf{v}_k^\theta \in \mathbb{R}^{N_\theta}$ and $\mathbf{v}_k^p \in \mathbb{R}^{N_p}$ are zero-mean GWN processes having covariance matrices $\mathbf{Q}^\theta \in \mathbb{R}^{N_\theta \times N_\theta}$ and $\mathbf{Q}^p \in \mathbb{R}^{N_p \times N_p}$, respectively. These stochastic processes are also advantageous for creating an augmented state vector, encompassing the states, parameters, and input forces, whereby their correlation can be considered. By doing so, the augmented state-space model can be expressed as

$$\mathbf{\xi}_k = f(\mathbf{\xi}_{k-1}) + \mathbf{v}_k^a \tag{9}$$

where $\mathbf{\xi}_k = [(\mathbf{z}_k)^T \; (\mathbf{\theta}_k)^T \; (\mathbf{p}_k)^T]^T$ is the augmented state vector consisting of $N_\xi = 2N_d + N_\theta + N_p$ unknown components; $\mathbf{v}_k^a = [(\mathbf{v}_k^z)^T \; (\mathbf{v}_k^\theta)^T \; (\mathbf{v}_k^p)^T]^T$ is the augmented process noise vector, described by a zero-mean GWN process with the covariance matrix $\mathbf{Q}^a = \text{block-diag}[\mathbf{Q}^z, \mathbf{Q}^\theta, \mathbf{Q}^p]$; for linear structures, the functional relationship $f(.)$ is expressed as

$$f(\mathbf{\xi}_{k-1}) = \mathbf{A}^a(\mathbf{\theta}_{k-1})\mathbf{\xi}_{k-1} \tag{10}$$

In this equation, $\mathbf{A}^a(\mathbf{\theta}_{k-1})$ is the augmented system matrix given by

$$\mathbf{A}^a(\mathbf{\theta}_{k-1}) = \begin{bmatrix} \mathbf{A}(\mathbf{\theta}_{k-1}) & \mathbf{0}_{2N_d \times N_\theta} & \mathbf{B}(\mathbf{\theta}_{k-1}) \\ \mathbf{0}_{N_\theta \times 2N_d} & \mathbf{I}_{N_\theta \times N_\theta} & \mathbf{0}_{N_\theta \times N_p} \\ \mathbf{0}_{N_p \times 2N_d} & \mathbf{0}_{N_p \times N_\theta} & \mathbf{I}_{N_p \times N_p} \end{bmatrix} \tag{11}$$

The state-space representation in Eq. (9) is linear with respect to the input forces, but the augmented process model is nonlinear with respect to the system state and the structural parameters. To circumvent this nonlinearity, a linearization with respect to $\mathbf{\xi}_{k-1}$ can be established:

$$\mathbf{\xi}_k = f(\hat{\mathbf{\xi}}_{k-1}) + \mathbf{F}_{k-1}^\xi \left(\mathbf{\xi}_{k-1} - \hat{\mathbf{\xi}}_{k-1}\right) + \mathbf{v}_k^a \tag{12}$$

where $\hat{\boldsymbol{\xi}}_{k-1}$ is the expansion point; $\mathbf{F}_{k-1}^{\xi} = \partial f(\boldsymbol{\xi}_{k-1})/\partial \boldsymbol{\xi}_{k-1}^{T}\big|_{\boldsymbol{\xi}_{k-1}=\hat{\boldsymbol{\xi}}_{k-1}}$ is the derivative of the process model with respect to the augmented state vector, evaluated at $\hat{\boldsymbol{\xi}}_{k-1}$ and determined from

$$\mathbf{F}_{k-1}^{\xi} = \mathbf{A}^{a}(\hat{\boldsymbol{\theta}}_{k-1}) + \frac{\partial(\mathbf{A}^{a}(\boldsymbol{\theta}_{k-1})\hat{\boldsymbol{\xi}}_{k-1})}{\partial \boldsymbol{\xi}_{k-1}^{T}}\bigg|_{\boldsymbol{\xi}_{k-1}=\hat{\boldsymbol{\xi}}_{k-1}} \quad (13)$$

This linearization has extensively been used in conjunction with the EKF to simplify state-space models. When the sampling interval and the variation in the structural parameters are relatively small, the linearization remains valid [47]. Later, we will also employ this relationship to derive the Bayesian formulation.

## 2.2. Observation Model

Let the measured quantities comprise strain, displacement, velocity, and acceleration responses. Then, the continuous-time observation model can be expressed as

$$\mathbf{d}(t) = \mathbf{G}_{c}(\boldsymbol{\theta}(t))\mathbf{z}(t) + \mathbf{J}_{c}\mathbf{p}(t) \quad (14)$$

and

$$\mathbf{d}(t) = [(\mathbf{S}_{\varepsilon}\boldsymbol{\varepsilon}(t))^{T} \ (\mathbf{S}_{d}\mathbf{x}(t))^{T} \ (\mathbf{S}_{v}\dot{\mathbf{x}}(t))^{T} \ (\mathbf{S}_{a}\ddot{\mathbf{x}}(t))^{T}]^{T} \quad (15)$$

$$\mathbf{G}_{c}(\boldsymbol{\theta}(t)) = \begin{bmatrix} \mathbf{S}_{\varepsilon}\mathbf{N}_{\varepsilon} & \mathbf{0}_{n_{\varepsilon}\times N_{d}} \\ \mathbf{S}_{d} & \mathbf{0}_{n_{d}\times N_{d}} \\ \mathbf{0}_{n_{v}\times N_{d}} & \mathbf{S}_{v} \\ -\mathbf{S}_{a}\mathbf{M}^{-1}\mathbf{K}(\boldsymbol{\theta}) & -\mathbf{S}_{a}\mathbf{M}^{-1}\mathbf{C}(\boldsymbol{\theta}) \end{bmatrix} \quad ; \quad \mathbf{J}_{c} = \begin{bmatrix} \mathbf{0}_{n_{\varepsilon}\times N_{p}} \\ \mathbf{0}_{n_{d}\times N_{p}} \\ \mathbf{0}_{n_{v}\times N_{p}} \\ \mathbf{S}_{a}\mathbf{M}^{-1}\mathbf{S}_{p} \end{bmatrix} \quad (16)$$

where $\mathbf{d}(t) \in \mathbb{R}^{N_{m}}$ denotes the measured response; $\boldsymbol{\varepsilon}(t) \in \mathbb{R}^{N_{\varepsilon}}$ denotes a vector of local strains comprising $N_{\varepsilon}$ elements; $\mathbf{S}_{\varepsilon} \in \mathbb{R}^{n_{\varepsilon}\times N_{\varepsilon}}$, $\mathbf{S}_{d} \in \mathbb{R}^{n_{d}\times N_{d}}$, $\mathbf{S}_{v} \in \mathbb{R}^{n_{v}\times N_{d}}$, and $\mathbf{S}_{a} \in \mathbb{R}^{n_{a}\times N_{d}}$ are known selection matrices, introduced for specifying the spatial distribution of strain, displacement, velocity, and acceleration sensors, respectively; $\mathbf{N}_{\varepsilon} \in \mathbb{R}^{N_{\varepsilon}\times N_{d}}$ is a known shape function matrix, mapping nodal displacement responses to local strains. Due to this definition, the sensor configuration consists of $n_{\varepsilon}$ strain, $n_{d}$ displacement, $n_{v}$ velocity, and $n_{a}$ acceleration responses, in total constituting $N_{m}$ response quantities. This continuous-time representation is beneficial for studying the observability conditions and deriving the discrete-time representation.

Without loss of generality, the sampling interval is considered the same as the process model. Let $D_{n} = \{\mathbf{d}_{k} = \mathbf{d}(t_{k}), k=1,2,...,n\}$ denote a sequence of response histories, where each individual sample of the measured response $\mathbf{d}_{k} \in \mathbb{R}^{N_{m}}$ corresponds to the time instant $t_{k} = k\Delta t$. Then, the discrete-time observation model is expressed as a function of the augmented state vector:

$$\mathbf{d}_{k} = \mathbf{G}(\boldsymbol{\theta}_{k})\boldsymbol{\xi}_{k} + \mathbf{w}_{k} \quad (17)$$

where $\mathbf{w}_{k} \in \mathbb{R}^{N_{m}}$ is the observation noise, considered to be GWN process with zero mean and $\mathbf{R} \in \mathbb{R}^{N_{m}\times N_{m}}$ covariance matrix; $\mathbf{G}(\boldsymbol{\theta}_{k}) \in \mathbb{R}^{N_{m}\times N_{\xi}}$ is the discrete-time observation matrix, given by

$$\mathbf{G}(\boldsymbol{\theta}_{k}) = [\mathbf{G}_{c}(\boldsymbol{\theta}_{k}) \quad \mathbf{0}_{N_{m}\times N_{\theta}} \quad \mathbf{J}_{c}] \quad (18)$$

The presence of low-frequency drifts is an obstacle for estimating input, state, and parameters, particularly when dealing with classical variants of Kalman filters [8]. This issue is particularly concerned when only acceleration measurements are accessible. In this regard, an effective strategy is to include pseudo-observations to impose prior information on unobserved quantities such that their low-frequency components stabilize and converge to bounded values. The core concept is to consider auxiliary variables, described by a zero-mean GWN with suitable variances. In the literature, displacement pseudo-observations have been used for reducing fictitious low-frequency drifts [10,48,49]. Recently, it has been shown that input pseudo-observations can help mitigate the drifts up to a great extent [45], as further promoted in this study.

Let $\mathbf{d}_k^{pd} \in \mathbb{R}^{N_{pd}}$ be a vector of input pseudo-observations. This vector can be expressed in terms of the augmented state vector as follows:

$$\mathbf{d}_k^{pd} = \mathbf{G}^{pd} \boldsymbol{\xi}_k + \mathbf{w}_k^{pd} \tag{19}$$

where $\mathbf{G}^{pd} = [\mathbf{0}_{N_{pd} \times 2N_d} \ \mathbf{0}_{N_{pd} \times N_\theta} \ \mathbf{I}_{N_{pd} \times N_p}]$ is a transformation matrix, mapping the state to the pseudo-data vector; $\mathbf{w}_k^{pd} \in \mathbb{R}^{N_{pd}}$ is a noise vector described by a zero-mean GWN process with $\mathbf{R}^{pd} \in \mathbb{R}^{N_{pd} \times N_{pd}}$ covariance matrix that should be updated by the data; $\mathbf{d}_k^{pd}$ is considered zero. By this definition, the pseudo-observations act as a zero-mean GWN process with $\mathbf{R}^{pd}$ unknown covariance matrix. Combining Eqs. (17) and (19) gives an augmented observation model as follows:

$$\mathbf{d}_k^a = h(\boldsymbol{\xi}_k) + \mathbf{w}_k^a \tag{20}$$

where $\mathbf{d}_k^a = [(\mathbf{d}_k)^T \ (\mathbf{d}_k^{pd})^T]^T$ is the augmented observation vector; $\mathbf{w}_k^a = [(\mathbf{w}_k)^T \ (\mathbf{w}_k^{pd})^T]^T$ is the augmented observation noise, assumed to be zero-mean GWN with the covariance matrix $\mathbf{R}^a = \text{block-diag}[\mathbf{R}, \mathbf{R}^{pd}]$; Given this noise setting, no cross-correlation is considered between the measured and dummy observations. We note that the functional relationship $h(.)$ is described as

$$h(\boldsymbol{\xi}_k) = \mathbf{G}^a(\boldsymbol{\theta}_k) \boldsymbol{\xi}_k \quad ; \quad \mathbf{G}^a(\boldsymbol{\theta}_k) = \begin{bmatrix} \mathbf{G}(\boldsymbol{\theta}_k) \\ \mathbf{G}^{pd} \end{bmatrix} \tag{21}$$

where $\mathbf{G}^a(\boldsymbol{\theta}_k) \in \mathbb{R}^{(N_m + N_{pd}) \times N_\xi}$ is the augmented observation matrix. The observation model is non-linear with respect to the augmented state vector, which can be approximated using a Taylor series expansion, giving:

$$\mathbf{d}_k^a = h(\hat{\boldsymbol{\xi}}_k) + \mathbf{H}_k^\xi (\boldsymbol{\xi}_k - \hat{\boldsymbol{\xi}}_k) + \mathbf{w}_k^a \tag{22}$$

where $\hat{\boldsymbol{\xi}}_k$ is the expansion point; $\mathbf{H}_k^\xi = \partial h(\boldsymbol{\xi}_k) / \partial \boldsymbol{\xi}_k^T \big|_{\boldsymbol{\xi}_k = \hat{\boldsymbol{\xi}}_k}$ is the derivative of the augmented observation model with respect to $\boldsymbol{\xi}_k$, calculated at $\hat{\boldsymbol{\xi}}_k$ and obtained as

$$\mathbf{H}_k^\xi = \mathbf{G}^a(\hat{\boldsymbol{\theta}}_k) + \frac{\partial (\mathbf{G}^a(\boldsymbol{\theta}_k) \hat{\boldsymbol{\xi}}_k)}{\partial \boldsymbol{\xi}_k^T} \bigg|_{\boldsymbol{\xi}_k = \hat{\boldsymbol{\xi}}_k} \tag{23}$$

Later, this linearization will be used in the derivation of Bayesian updating formulation.

### 2.3. Virtual Sensing Equations

Virtual sensing uses the estimates of the augmented state vector for predicting unmeasured responses, including stress, strain, displacement, velocity, and acceleration response histories. For this purpose, we should characterize the relationship between the augmented state and these response quantities, which is given by

$$\mathbf{d}_k^e = \mathbf{G}^e(\boldsymbol{\theta}_k) \boldsymbol{\xi}_k \tag{24}$$

where $\mathbf{d}_k^e \in \mathbb{R}^{N_e}$ comprises virtual responses, and $\mathbf{G}^e(\boldsymbol{\theta}_k) \in \mathbb{R}^{N_e \times N_\xi}$ is the transformation matrix. When full-field response reconstruction is desired, the matrix $\mathbf{G}^e(\boldsymbol{\theta}_k)$ is calculated as

$$\mathbf{G}^e(\boldsymbol{\theta}_k) = \begin{bmatrix} \mathbf{B}_\sigma & \mathbf{0}_{N_\sigma \times N_d} & \mathbf{0}_{N_\sigma \times N_\theta} & \mathbf{0}_{N_\sigma \times N_p} \\ \mathbf{N}_\varepsilon & \mathbf{0}_{N_\varepsilon \times N_d} & \mathbf{0}_{N_\varepsilon \times N_\theta} & \mathbf{0}_{N_\varepsilon \times N_p} \\ \mathbf{I}_{N_d \times N_d} & \mathbf{0}_{N_d \times N_d} & \mathbf{0}_{N_d \times N_\theta} & \mathbf{0}_{N_d \times N_p} \\ \mathbf{0}_{N_d \times N_d} & \mathbf{I}_{N_d \times N_d} & \mathbf{0}_{N_d \times N_\theta} & \mathbf{0}_{N_d \times N_p} \\ -\mathbf{M}^{-1}\mathbf{K}(\boldsymbol{\theta}_k) & -\mathbf{M}^{-1}\mathbf{C}(\boldsymbol{\theta}_k) & \mathbf{0}_{N_d \times N_\theta} & \mathbf{M}^{-1}\mathbf{S}_p \end{bmatrix} \quad (25)$$

where $\mathbf{B}_\sigma \in \mathbb{R}^{N_\sigma \times N_d}$ is a known matrix mapping displacements into stress responses. When the structural parameters are unknown, this virtual sensing equation is also nonlinear with respect to the augmented state. A convenient approach is to linearize Eq. (24) by writing the first-order Taylor series expansion around $\hat{\boldsymbol{\xi}}_k$, which gives:

$$\mathbf{d}_k^e = \mathbf{G}^e(\hat{\boldsymbol{\theta}}_k)\hat{\boldsymbol{\xi}}_k + \left.\frac{\partial (\mathbf{G}^e(\boldsymbol{\theta}_k)\hat{\boldsymbol{\xi}}_k)}{\partial \boldsymbol{\xi}_k^T}\right|_{\boldsymbol{\xi}_k=\hat{\boldsymbol{\xi}}_k} (\boldsymbol{\xi}_k - \hat{\boldsymbol{\xi}}_k) \quad (26)$$

This approximation will later be used to characterize the uncertainty associated with the virtually-sensed responses.

### 2.4. Theoretical Observability

The observability of the input, states, and parameters is a theoretical tool for testing whether a particular sensor configuration suffices. In this respect, Lie-derivatives-based methods have shown efficiency and robustness [50–53]. To apply them, the Observability Rank Criterion (ORC) should be derived mathematically based on the continuous-time process and measurement models. Fortunately, the linearity of the stiffness and damping matrices with respect to the structural parameters allows for obtaining an explicit formulation for the ORC matrix. According to the detailed derivation provided in Appendix (A), the ORC matrix is calculated as

$$d\Omega_k = \begin{bmatrix} \mathbf{G}_0 & \mathbf{H} & \mathbf{J}_c & \mathbf{0} & \mathbf{0} & \mathbf{0} & \mathbf{0} \\ \mathbf{G}_0\mathbf{A}_0 & \mathbf{H}+\mathbf{G}_0\mathbf{C} & \mathbf{G}_0\mathbf{B}_c & \mathbf{J}_c & \mathbf{0} & \mathbf{0} & \mathbf{0} \\ \mathbf{G}_0\mathbf{A}_0^2 & \mathbf{H}+\mathbf{G}_0\mathbf{C}+\mathbf{G}_0\mathbf{A}_0\mathbf{C} & \mathbf{G}_0\mathbf{A}_0\mathbf{B}_c & \mathbf{G}_0\mathbf{B}_c & \mathbf{J}_c & \mathbf{0} & \mathbf{0} \\ \vdots & \vdots & \vdots & \ddots & \ddots & \ddots & \vdots \\ \mathbf{G}_0\mathbf{A}_0^k & \mathbf{H}+\mathbf{G}_0\mathbf{C}+\sum_{j=1}^{k-1}\mathbf{G}_0\mathbf{A}_0^j\mathbf{C} & \mathbf{G}_0\mathbf{A}_0^{k-1}\mathbf{B}_c & \cdots & \mathbf{G}_0\mathbf{A}_0\mathbf{B}_c & \mathbf{G}_0\mathbf{B}_c & \mathbf{J}_c \end{bmatrix} \quad (27)$$

$$\underbrace{\phantom{XXXX}}_{\mathcal{O}_k} \underbrace{\phantom{XXXXXXXXXXX}}_{\mathcal{G}_k} \underbrace{\phantom{XX}}_{\mathbb{H}_k}$$

where $d\Omega_k = [\mathcal{O}_k \; \mathcal{G}_k \; \mathbb{H}_k]$ denotes the ORC matrix, representing the so-called $k^{\text{th}}$-row observability of the continuous-time vector $\boldsymbol{\eta}(t) = [(\boldsymbol{\xi}(t))^T \; (\dot{\mathbf{p}}(t))^T \; ... \; (\mathbf{p}^{(k)}(t))^T]^T$, which comprises the augmented state and higher-order derivatives of the input forces up to $k^{\text{th}}$ order; $\mathbf{A}_0 \triangleq \mathbf{A}_c(\boldsymbol{\theta})|_{\boldsymbol{\theta}=\mathbf{0}}$ and $\mathbf{G}_0 \triangleq \mathbf{G}_c(\boldsymbol{\theta})|_{\boldsymbol{\theta}=\mathbf{0}}$ are the system and state-to-observation matrices evaluated at $\boldsymbol{\theta}(t)|_{t=0} = \mathbf{0}$, respectively; $\mathbf{C} = [\mathbf{A}_1\mathbf{z}_0 \; \cdots \; \mathbf{A}_{N_\theta}\mathbf{z}_0]$ and $\mathbf{H} = [\mathbf{G}_1\mathbf{z}_0 \; \cdots \; \mathbf{G}_{N_\theta}\mathbf{z}_0]$ are known matrices, whose columns are filled by the derivatives of $\mathbf{A}_c(\boldsymbol{\theta}(t))\mathbf{z}_0$ and $\mathbf{G}_c(\boldsymbol{\theta}(t))\mathbf{z}_0$ with respect to individual elements of $\boldsymbol{\theta}$, respectively (see Eqs. (A8-A9) in Appendix A); $\mathbf{A}_i$ and $\mathbf{G}_i$ are obtained from the mathematical expressions of $\mathbf{A}_c(\boldsymbol{\theta}(t))$ and $\mathbf{G}_c(\boldsymbol{\theta}(t))$, wherein $\boldsymbol{\theta}$ is replaced by a single-entry vector whose $i^{\text{th}}$ element is one (see Eqs. (A4-A5) in Appendix A); $\mathbf{z}_0$ is an arbitrary choice of the state vector $\mathbf{z}(t)$ used for the linearization.

The vector $\boldsymbol{\eta}(t)$ is $k^{\text{th}}$-row observable if the rank of $d\Omega_k$ is equal to $2N_d + N_\theta + (k+1)N_p$. Satisfying this condition implies that the coupled measurement and process model is locally weakly observable [50]. In situations where this condition is not satisfied and the vector $\boldsymbol{\eta}(t)$ is not fully observable, the $k^{\text{th}}$-row observability can still be investigated for each component of $\mathbf{\eta}(t)$. By definition, the $m^{\text{th}}$ element of the state vector is $k^{\text{th}}$-row observable if removing the $m^{\text{th}}$ column of $d\Omega_k$ reduces the rank by one. Otherwise, the investigated component of $\mathbf{\eta}(t)$ is $k^{\text{th}}$-row unobservable. It should be noticed that if a quantity is $k^{\text{th}}$-row observable, it is also higher-row observable, e.g., $(k+1)^{\text{th}}$-row observable [51].

The rank condition of $d\Omega_k$ can be discussed in terms of the rank of the three block columns $\mathcal{O}_k$, $\mathcal{G}_k$, and $\mathcal{H}_k$, which govern the observability of $\mathbf{\eta}(t)$. The rank condition of $\mathcal{O}_k$ can be understood as the observability condition in the absence of unknown parameters and input forces. This result implies that the conventional observability condition, $Rank(\mathcal{O}_k) = 2N_d$, is an enforced requisite for the observability in the presence of unknown parameters and input. When unknown parameters also exist, the rank condition transforms into $Rank([\mathcal{O}_k \quad \mathcal{G}_k]) = 2N_d + N_\theta$. The latter condition is more general than the former, so it is sufficient to satisfy only the second condition to ensure that the dynamical states and parameters are both observable. Regarding the input forces, as proven in [51], the presence of unknown input forces dictates satisfying $Rank(\mathcal{H}_{k-1}) = Rank(\mathcal{H}_k) - N_p$. Therefore, the $k^{\text{th}}$-row observability of $\mathbf{\eta}(t)$ can be generalized as satisfying:

$$\begin{cases} Rank([\mathcal{O}_k \quad \mathcal{G}_k]) = 2N_d + N_\theta \\ Rank(\mathcal{H}_{k-1}) = Rank(\mathcal{H}_k) - N_p \end{cases} \tag{28}$$

In the illustrative examples, we will elaborate on how this rank requirement should be checked for presumed sensor setups.

## 3. Bayesian Formulation

The primary goal of this study is to establish a probabilistic framework to tackle the coupled estimation problem while quantifying the uncertainties through updated noise characteristics. In the Bayesian sense, this approach requires the calculation of the posterior probability density function (PDF) of the latent states and noise covariance matrices.

We remind that $D_n = \{\mathbf{d}_k, k = 1, 2, ..., n\}$ is a sequence of response histories, where the subscript $k$ represents the discrete time $t_k = k\Delta t$. The observation model described by Eq. (20) relates the augmented observed response ($\mathbf{d}_k^a$) to the augmented state vector ($\mathbf{\xi}_k$) and the measurement noise covariance matrix ($\mathbf{R}^a$). This relationship implies that the full dataset likelihood function can be expressed as the multiplication of $n$ independent Gaussian distributions:

$$p(D_n \mid \{\mathbf{\xi}_k\}_{k=1}^n, \mathbf{Q}^a, \mathbf{R}^a) = \prod_{k=1}^n N(\mathbf{d}_k^a \mid h(\mathbf{\xi}_k), \mathbf{R}^a) \tag{29}$$

Additionally, the joint prior PDF can be simplified based on the process model described by Eq. (9), which leads to

$$p(\{\mathbf{\xi}_k\}_{k=0}^n, \mathbf{Q}^a, \mathbf{R}^a) = p(\mathbf{Q}^a, \mathbf{R}^a) p(\mathbf{\xi}_0) \prod_{k=1}^n p(\mathbf{\xi}_k \mid \mathbf{\xi}_{k-1}, \mathbf{Q}^a) \tag{30}$$

where $p(\{\mathbf{\xi}_k\}_{k=0}^n, \mathbf{Q}^a, \mathbf{R}^a)$ is the joint prior PDF; $p(\mathbf{Q}^a, \mathbf{R}^a)$ is the prior PDF of the noise covariance matrices, considered to be a uniform distribution; $p(\mathbf{\xi}_0)$ is the prior PDF of the augmented state at time $t = 0$, considered to be Gaussian having mean vector ($\mathbf{\mu}_{\xi_0}$) and covariance matrix ($\mathbf{P}_{\xi_0}$); $p(\mathbf{\xi}_k \mid \mathbf{\xi}_{k-1}, \mathbf{Q}^a)$ is the transitional distribution of the augmented state, described by Eq. (9) as $N(\mathbf{\xi}_k \mid f(\mathbf{\xi}_{k-1}), \mathbf{Q}^a)$. Therefore, the joint prior PDF can be simplified into

$$p(\{\mathbf{\xi}_k\}_{k=0}^n, \mathbf{\mu}_{\xi_0}, \mathbf{P}_{\xi_0}, \mathbf{Q}^a, \mathbf{R}^a) \propto N(\mathbf{\xi}_0 \mid \mathbf{\mu}_{\xi_0}, \mathbf{P}_{\xi_0}) \prod_{k=1}^n N(\mathbf{\xi}_k \mid f(\mathbf{\xi}_{k-1}), \mathbf{Q}^a) \tag{31}$$

Here, the parameters $\boldsymbol{\mu}_{\xi_0}$ and $\mathbf{P}_{\xi_0}$ are added to the joint prior PDF since they should also be updated based on the data. Then, the joint posterior PDF can be calculated by the Bayes' rule:

$$p(\{\boldsymbol{\xi}_k\}_{k=0}^n, \boldsymbol{\mu}_{\xi_0}, \mathbf{P}_{\xi_0}, \mathbf{Q}^a, \mathbf{R}^a \mid D_n) \propto N(\boldsymbol{\xi}_0 \mid \boldsymbol{\mu}_{\xi_0}, \mathbf{P}_{\xi_0}) \prod_{k=1}^n N(\mathbf{d}_k^a \mid h(\boldsymbol{\xi}_k), \mathbf{R}^a) N(\boldsymbol{\xi}_k \mid f(\boldsymbol{\xi}_{k-1}), \mathbf{Q}^a) \qquad (32)$$

where $p(\{\boldsymbol{\xi}_k\}_{k=0}^n, \boldsymbol{\mu}_{\xi_0}, \mathbf{P}_{\xi_0}, \mathbf{Q}^a, \mathbf{R}^a \mid D_n)$ is the joint posterior PDF. Although this PDF looks simple, the computation is non-trivial and challenging, especially when the system is partially known and the measurements are spatially sparse. In addition, the Markovian dependence of the latent states increases the complexity of calculations in the sense that all states have to be optimized simultaneously.

To better understand how the parameters and observed data are related, a graphical representation of the probabilistic model is displayed in Fig. 1. The arrows show the conditional dependence between the parameters. As shown, the latent state $\boldsymbol{\xi}_k$ depends on all preceding states, as well as the noise parameters. When the noise characteristics and the initial state parameters are known, the identification of the latent states can be straightforward using filter-type techniques. This understanding aids to implement an efficient computation method, discussed in the next section.

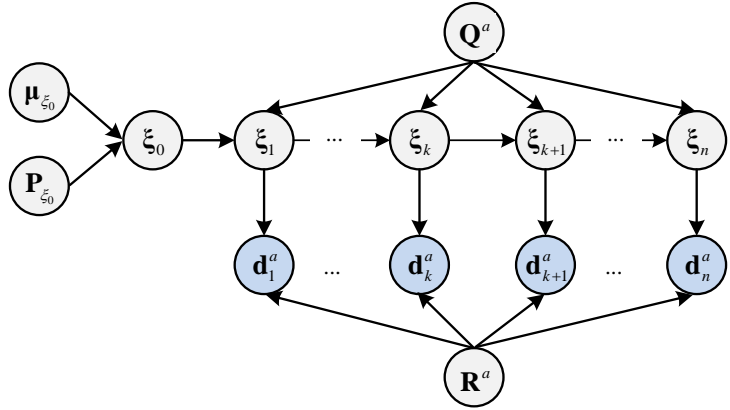

**Fig. 1**. A hierarchical representation of the proposed probabilistic model and its underlying parameters

## 4. Derivation of the Computation Approach

In the context of Bayesian methods, the most probable values (MPVs) can be calculated through the direct maximization of the joint posterior PDF with respect to the unknown parameters. However, this approach is not practicable for the posterior PDF obtained in Eq. (32). Thus, it is preferred to maximize a better posed substitute function whose maximum in-limit converges to those of the posterior PDF. The EM algorithm is a good example for this type of strategy, employed herein. Specifically, the $j^{\text{th}}$ iteration of the proposed algorithm requires taking two main steps, summarized below:

E-Step: Calculate $L(\Phi \mid \hat{\Phi}^{(j-1)}) = \mathbb{E}_{\Xi_n}[\ln p(\Xi_n, \Phi \mid D_n)]$ \hfill (33)

M-Step: Maximize $L(\Phi \mid \hat{\Phi}^{(j-1)})$ with respect to $\Phi$ \hfill (34)

where the E-Step and M-Step refer to "Expectation" and "Maximization," respectively; the set $\Phi = \{\boldsymbol{\mu}_{\xi_0}, \mathbf{P}_{\xi_0}, \mathbf{Q}^a, \mathbf{R}^a\}$ comprises the unknown hyper-parameters, treated as being invariant across data points; $\hat{\Phi}^{(j-1)}$ is the optimal value of $\Phi$ obtained at the $(j-1)^{\text{th}}$ iteration of the EM algorithm; the expectation $\mathbb{E}_{\Xi_n}[.]$ should be taken with respect to the full set of latent states $\Xi_n = \{\boldsymbol{\xi}_k\}_{k=0}^n$; $L(\Phi \mid \hat{\Phi}^{(j-1)})$ is the substitute function, which is indirectly related to the natural logarithm of the following posterior PDF:

$$p(\Xi_n, \Phi \mid D_n) \propto N(\boldsymbol{\xi}_0 \mid \boldsymbol{\mu}_{\xi_0}, \mathbf{P}_{\xi_0}) \prod_{k=1}^n N(\mathbf{d}_k^a \mid h(\boldsymbol{\xi}_k), \mathbf{R}^a) N(\boldsymbol{\xi}_k \mid f(\boldsymbol{\xi}_{k-1}), \mathbf{Q}^a) \qquad (35)$$

From this equation, one can directly obtain:

$$\ln p(\Xi_n, \Phi | D_n) = \ln N(\xi_0 | \mu_{\xi_0}, P_{\xi_0}) + \sum_{k=1}^{n} \ln N(\mathbf{d}_k^a | h(\xi_k), \mathbf{R}^a) + \sum_{k=1}^{n} \ln N(\xi_k | f(\xi_{k-1}), \mathbf{Q}^a) + cte.$$

$$= -\frac{1}{2}|\mathbf{P}_{\xi_0}| - \frac{1}{2}(\xi_0 - \mu_{\xi_0})^T \mathbf{P}_{\xi_0}^{-1}(\xi_0 - \mu_{\xi_0})$$

$$-\frac{n}{2}|\mathbf{R}^a| - \frac{1}{2}\sum_{k=1}^{n}(\mathbf{d}_k^a - h(\xi_k))^T (\mathbf{R}^a)^{-1}(\mathbf{d}_k^a - h(\xi_k))$$

$$-\frac{n}{2}|\mathbf{Q}^a| - \frac{1}{2}\sum_{k=1}^{n}(\xi_k - f(\xi_{k-1}))^T (\mathbf{Q}^a)^{-1}(\xi_k - f(\xi_{k-1})) + cte.$$

(36)

By taking expectation with respect to the latent states, the substitute function in Eq. (33) will be obtained as

$$L(\Phi | \hat{\Phi}^{(j-1)}) = -\frac{n}{2}\ln|\mathbf{R}^a| - \frac{n}{2}\ln|\mathbf{Q}^a| - \frac{1}{2}\ln|\mathbf{P}_{\xi_0}| + \frac{1}{2}tr\left((\mathbf{P}_{\xi_0})^{-1}\mathbb{E}_{\xi_0}[(\xi_0 - \mu_{\xi_0})(\xi_0 - \mu_{\xi_0})^T]\right)$$

$$+\frac{1}{2}tr\left((\mathbf{R}^a)^{-1}\sum_{k=1}^{n}\mathbb{E}_{\xi_k}[(\mathbf{d}_k^a - h(\xi_k))(\mathbf{d}_k^a - h(\xi_k))^T] + (\mathbf{Q}^a)^{-1}\sum_{k=1}^{n}\mathbb{E}_{\xi_k,\xi_{k-1}}[(\xi_k - f(\xi_{k-1}))(\xi_k - f(\xi_{k-1}))^T]\right)$$

(37)

In this equation, the Mahalanobis distance is written in its alternative form given as $\mathbf{x}^T \mathbf{A} \mathbf{x} = tr(\mathbf{A}\mathbf{x}\mathbf{x}^T)$.

From Eqs. (33) and (37), it can be realized that the E-Step requires computing second-moment statistical information of the full set of latent states ($\Xi_n$) conditional on the prior estimation of $\Phi$ denoted by $\hat{\Phi}^{(j-1)}$. Then, the surrogate function $L(\Phi | \hat{\Phi}^{(j-1)})$ should be maximized for $\Phi$ to update the estimations. These steps should be repeated until the convergence is attained.

It should be noted that the EM algorithm provides a monotonous convergence to MPVs under some normality conditions [43,54]. However, similar to any other optimization algorithm, a suitable set of starting values can greatly accelerate the convergence rate while avoiding undesirable maxima. In the next section, the derivation of the EM algorithm is explained, and an efficient initializer is introduced in Appendix (B).

### 4.1. Calculation of the Expectations (E-Step)

Evaluating the function $L(\Phi | \hat{\Phi}^{(j-1)})$ requires the statistical moments of the latent states, which include their mean, covariance, and cross-covariance. This information should be determined by calculating the posterior PDF of the latent states conditional on $\Phi$ and $D_n$, i.e., $p(\Xi_n | \Phi, D_n)$. This distribution can be derived through a backward-forward Bayesian smoother.

The forward Bayesian estimator enables updating the states through a forward recursive formulation. The derivation of this estimator is convenient through mathematical induction. Let the induction base hold as $p(\xi_0 | \Phi) = N(\xi_0 | \mu_{\xi_0}, \mathbf{P}_{\xi_0})$. Then, the hypothesis is that the posterior PDF of the state $\xi_{k-1}$ conditional on $\Phi$ and $D_{k-1}$ is given by

$$p(\xi_{k-1} | \Phi, D_{k-1}) = N(\xi_{k-1} | \xi_{k-1|k-1}, \mathbf{P}_{k-1|k-1}^{\xi}) \quad (38)$$

where $\xi_{k-1|k-1} \in \mathbb{R}^{N_\xi}$ and $\mathbf{P}_{k-1|k-1}^{\xi} \in \mathbb{R}^{N_\xi \times N_\xi}$ are respectively the mean vector and covariance matrix, obtained at the $(k-1)^{th}$ step of the forward estimator. It is desired to prove that a similar distribution holds for the state $\xi_k$ conditional on $\Phi$ and $D_k$, whose underlying parameters should be calculated.

For this purpose, it is first assumed that the transitional distribution of $\xi_k$ conditional on $\Phi$ and $D_{k-1}$ is constructed based on the process model. Then, the linearized approximation of the process model given by Eq. (12) is used, which yields [55]:

$$p(\xi_k | \Phi, D_{k-1}) = N(\xi_k | \xi_{k|k-1}, \mathbf{P}_{k|k-1}^{\xi}) \quad (39)$$

Here, the predictive mean vector $\xi_{k|k-1} \in \mathbb{R}^{N_\xi}$ and covariance matrix $\mathbf{P}_{k|k-1}^{\xi} \in \mathbb{R}^{N_\xi \times N_\xi}$ are calculated as

$$\xi_{k|k-1} = f(\xi_{k-1|k-1}) \tag{40}$$

$$\mathbf{P}_{k|k-1}^{\xi} = \mathbf{F}_{k-1|k-1}^{\xi} \mathbf{P}_{k-1|k-1}^{\xi} (\mathbf{F}_{k-1|k-1}^{\xi})^T + \mathbf{Q}^a \tag{41}$$

where $\mathbf{F}_{k-1|k-1}^{\xi} = \partial f(\xi_{k-1}) / \partial \xi_{k-1}^T \big|_{\xi_{k-1}=\xi_{k-1|k-1}}$ is given by Eq. (13). Given this transitional distribution, the Bayes' rule can be used to update the next increment of the state vector $\xi_k$ by incorporating the observation $\mathbf{d}_k^a$. This procedure is easily accomplished using the linearized observation model of Eq. (22), which leads to the conventional EKF formulation [55]:

$$p(\xi_k | \Phi, D_k) = N(\xi_k | \xi_{k|k}, \mathbf{P}_{k|k}^{\xi}) \tag{42}$$

Here, the mean and covariance are given as

$$\begin{aligned}
\mathbf{G}_k^{\xi} &= \mathbf{P}_{k|k-1}^{\xi} (\mathbf{H}_{k|k-1}^{\xi})^T (\mathbf{R}^a + \mathbf{H}_{k|k-1}^{\xi} \mathbf{P}_{k|k-1}^{\xi} (\mathbf{H}_{k|k-1}^{\xi})^T)^{-1} \\
\xi_{k|k} &= \xi_{k|k-1} + \mathbf{G}_k^{\xi} (\mathbf{d}_k^a - h(\xi_{k|k-1})) \\
\mathbf{P}_{k|k}^{\xi} &= \mathbf{P}_{k|k-1}^{\xi} - \mathbf{G}_k^{\xi} \mathbf{H}_{k|k-1}^{\xi} \mathbf{P}_{k|k-1}^{\xi}
\end{aligned} \tag{43}$$

where $\mathbf{G}_k^{\xi} \in \mathbb{R}^{N_{\xi} \times (N_m + N_{pd})}$ is the filter gain matrix that controls the transmission of information from the observed quantities to the estimates, and $\mathbf{H}_{k|k-1}^{\xi} = \partial h(\xi_k) / \partial \xi_k^T \big|_{\xi_k = \xi_{k|k-1}} \in \mathbb{R}^{(N_m + N_{pd}) \times N_{\xi}}$ is given by Eq. (24) and evaluated at $\xi_{k|k-1}$. The results obtained here complete the inductive step, confirming that the updated PDF of $\xi_k$ conditional on $D_k$ is also a Gaussian PDF. Thus, the recursive calculation of the mean and covariance of the latent states would suffice to characterize this PDF.

The EKF-based estimates of the states can be improved based on future observations using smoothing techniques, for which a few Bayesian strategies are available, including fixed-lag and fixed-interval approaches [55,56]. A special case of the latter approach is fixed-point smoothing method, whereby $\xi_k$ is re-estimated using $D_{k+1} = D_k \cup \mathbf{d}_{k+1}^a$ rather than $D_n$. Such a smoothing strategy results in the smoothed PDF of $\xi_k$, given by [55]

$$p(\xi_k | \Phi, D_n) \approx p(\xi_k | \Phi, D_{k+1}) = N(\xi_k | \xi_{k|k+1}, \mathbf{P}_{k|k+1}^{\xi}) \tag{44}$$

In this equation, $\xi_{k|k+1} \in \mathbb{R}^{N_{\xi}}$ and $\mathbf{P}_{k|k+1}^{\xi} \in \mathbb{R}^{N_{\xi} \times N_{\xi}}$ are respectively the smoothed mean vector and covariance matrix, calculated from

$$\begin{aligned}
\mathbf{L}_k^{\xi} &= \mathbf{P}_{k|k}^{\xi} (\mathbf{F}_k^{\xi})^T (\mathbf{P}_{k+1|k}^{\xi})^{-1} \\
\xi_{k|k+1} &= \xi_{k|k} + \mathbf{L}_k^{\xi} (\xi_{k+1|k+1} - \xi_{k+1|k}) \\
\mathbf{P}_{k|k+1}^{\xi} &= \mathbf{P}_{k|k}^{\xi} + \mathbf{L}_k^{\xi} (\mathbf{P}_{k+1|k+1}^{\xi} - \mathbf{P}_{k+1|k}^{\xi})(\mathbf{L}_k^{\xi})^T
\end{aligned} \tag{45}$$

where $\mathbf{L}_k^{\xi} \in \mathbb{R}^{N_{\xi} \times N_{\xi}}$ is the smoothing gain matrix, which corrects the mean and covariance of $\xi_k$. In the latest set of equations, the estimations $\xi_{k+1|k}$, $\xi_{k+1|k+1}$, $\mathbf{P}_{k+1|k}^{\xi}$, and $\mathbf{P}_{k+1|k+1}^{\xi}$ are required. Therefore, it is reasonable to perform this backward smoothing after running the EKF for $t_{k+1} = (k+1)\Delta t$. It should also be noticed that the cross covariance of $\xi_{k+1}$ and $\xi_k$ is calculated as $\mathbf{L}_k^{\xi} \mathbf{P}_{k|k+1}^{\xi}$ [55].

Based on the smoothed estimations of the latent states, the expectations encountered in Eq. (37) can be calculated using the first-order Taylor series expansions acquired earlier. By doing so, we get:

$$\mathbb{E}_{\xi_k}[(\mathbf{d}_k^a - h(\xi_k))(\mathbf{d}_k^a - h(\xi_k))^T] = (\mathbf{d}_k^a - h(\xi_{k|k+1}))(\mathbf{d}_k^a - h(\xi_{k|k+1}))^T + \mathbf{H}_{k|k+1}^{\xi} \mathbf{P}_{k|k+1}^{\xi} (\mathbf{H}_{k|k+1}^{\xi})^T \tag{46}$$

$$\begin{aligned}
\mathbb{E}_{\xi_k, \xi_{k-1}}[(\xi_k - f(\xi_{k-1}))(\xi_k - f(\xi_{k-1}))^T] &= (\xi_{k|k+1} - f(\xi_{k-1|k}))(\xi_{k|k+1} - f(\xi_{k-1|k}))^T \\
&+ \mathbf{P}_{k|k+1}^{\xi} + \mathbf{F}_{k-1|k}^{\xi} \mathbf{P}_{k-1|k}^{\xi} (\mathbf{F}_{k-1|k}^{\xi})^T - \mathbf{F}_{k-1|k}^{\xi} \mathbf{L}_{k-1}^{\xi} \mathbf{P}_{k|k}^{\xi} - \mathbf{P}_{k|k}^{\xi} (\mathbf{L}_{k-1}^{\xi})^T (\mathbf{F}_{k-1|k}^{\xi})^T
\end{aligned} \tag{47}$$

Substituting these equations into Eq. (37) completes the E-Step of the algorithm. Before moving to the M-Step, we note that the Bayesian smoothing uses the information before and after the current time instant to deliver a superior estimation of the states, which enhances the estimates' stability and mitigate noise disturbances.

### 4.2. Maximization of the Surrogate Function (M-Step)

In this section, the noise covariance matrices, as well as the initial state parameters are updated using the statistical information of the state vectors obtained in the E-Step. We first note that the derivatives of the function in Eq. (37) with respect to the components of $\Phi$ can be calculated as

$$\frac{\partial L(\Phi \mid \hat{\Phi}^{(j-1)})}{\partial \mathbf{\mu}_{\xi_0}} = (\mathbf{P}_{\xi_0})^{-1} \mathbb{E}_{\xi_0}[\xi_0 - \mathbf{\mu}_{\xi_0}] \tag{48}$$

$$\frac{\partial L(\Phi \mid \hat{\Phi}^{(j-1)})}{\partial \mathbf{P}_{\xi_0}} = -\frac{1}{2}(\mathbf{P}_{\xi_0})^{-1} + \frac{1}{2}(\mathbf{P}_{\xi_0})^{-1} \mathbb{E}_{\xi_0}[(\xi_0 - \mathbf{\mu}_{\xi_0})(\xi_0 - \mathbf{\mu}_{\xi_0})^T](\mathbf{P}_{\xi_0})^{-1} \tag{49}$$

$$\frac{\partial L(\Phi \mid \hat{\Phi}^{(j-1)})}{\partial \mathbf{R}^a} = -\frac{n}{2}(\mathbf{R}^a)^{-1} + \frac{1}{2}(\mathbf{R}^a)^{-1}\left[\sum_{k=1}^{n}\mathbb{E}_{\xi_k}[(\mathbf{d}_k^a - h(\xi_k))(\mathbf{d}_k^a - h(\xi_k))^T]\right](\mathbf{R}^a)^{-1} \tag{50}$$

$$\frac{\partial L(\Phi \mid \hat{\Phi}^{(j-1)})}{\partial \mathbf{Q}^a} = -\frac{n}{2}(\mathbf{Q}^a)^{-1} + \frac{1}{2}(\mathbf{Q}^a)^{-1}\left[\sum_{k=1}^{n}\mathbb{E}_{\xi_k,\xi_{k-1}}[(\xi_k - f(\xi_{k-1}))(\xi_k - f(\xi_{k-1}))^T]\right](\mathbf{Q}^a)^{-1} \tag{51}$$

where the expectations should be replaced from Eqs. (46-47). Setting these partial derivatives equal to zero will provide:

$$\hat{\mathbf{\mu}}_{\xi_0} = \xi_{0|1} \tag{52}$$

$$\hat{\mathbf{P}}_{\xi_0} = \mathbf{P}_{0|1}^{\xi} \tag{53}$$

$$\hat{\mathbf{R}}^a = \frac{1}{n}\sum_{k=1}^{n}\left[(\mathbf{d}_k^a - h(\xi_{k|k+1}))(\mathbf{d}_k^a - h(\xi_{k|k+1}))^T + \mathbf{H}_{k|k+1}^{\xi}\mathbf{P}_{k|k+1}^{\xi}(\mathbf{H}_{k|k+1}^{\xi})^T\right] \tag{54}$$

$$\hat{\mathbf{Q}}^a = \frac{1}{n}\sum_{k=1}^{n}\left[(\xi_{k|k+1} - f(\xi_{k-1|k}))(\xi_{k|k+1} - f(\xi_{k-1|k}))^T\right]$$
$$+ \frac{1}{n}\sum_{k=1}^{n}\left[\mathbf{P}_{k|k+1}^{\xi} + \mathbf{F}_{k-1|k}^{\xi}\mathbf{P}_{k-1|k}^{\xi}(\mathbf{F}_{k-1|k}^{\xi})^T - \mathbf{F}_{k-1|k}^{\xi}\mathbf{L}_{k-1}^{\xi}\mathbf{P}_{k-1|k}^{\xi} - \mathbf{P}_{k-1|k}^{\xi}(\mathbf{L}_{k-1}^{\xi})^T(\mathbf{F}_{k-1|k}^{\xi})^T\right] \tag{55}$$

where over-hat denotes the updated estimates of the parameters. The covariance matrices calculated herein are essentially positive semi-definite, so there is no need to consider any additional constraints. Moreover, there is no need to perform brute-force optimization, as required in [40].

### 5. Computational Algorithm

A summary of the BEM is presented in Algorithm 1, which consists of three main stages, as separated with the horizontal lines. In the first stage, the initial state, as well as the noise parameters should be introduced. For the structural parameters, a rough initial estimation would suffice, which can set to zero if a parameterization like Eqs. (2-3) is applied. The estimation of the state and input forces at $t = 0$ can be set to zero. Initial choices of the noise covariance matrices are also required, which involves specifying:

- Process noise covariance matrix ($\mathbf{Q}^a$) comprising:
    - State noise covariance matrix ($\mathbf{Q}^z$)
    - Input noise covariance matrix ($\mathbf{Q}^p$)
    - Parameter noise covariance matrix ($\mathbf{Q}^\theta$)
- Observation noise covariance matrix ($\mathbf{R}^a$) comprising:
    - Real data covariance matrix ($\mathbf{R}$)
    - Pseudo data covariance matrix ($\mathbf{R}^{pd}$)

The initial choice of $\mathbf{Q}^\theta$ can be a diagonal matrix with small entries, e.g., $10^{-8}\mathbf{I}_{N_\theta}$. For other noise covariance matrices ($\mathbf{Q}^z$ and $\mathbf{Q}^p$), Algorithm 2 of Appendix B can be employed. The Mean-Square (MS) of the measured quantities can provide a good starting value for $\mathbf{R}$, giving a diagonal matrix with elements $\mathbf{R}_{(i,i)} = \gamma \times \mathrm{MS}(d_i)$, where the coefficient $\gamma$ can be adjusted by the user based on the level of noise and modeling errors. The initial value of $\mathbf{R}^{pd}$ can be proportional to $\mathbf{Q}^p$. Note that these starting values are required to initiate Algorithm 1, and they will be updated iteratively based on the data.

Having introduced the initial values, in the second stage, the while-loop iteratively estimates the augmented state vector through the EKF and the fixed-point smoother and updates the noise covariance matrices. This procedure is repeated until the increase in the substitute function has reached a steady regime, satisfying a predefined tolerance. Additionally, this convergence criterion should be checked along with a cap on the maximum number of iterations to prevent an unnecessarily large number of trials. Once the convergence is achieved, the estimated states can be used to create virtual response estimates, as outlined in the third stage. Based on Eq. (26), the posterior predictive PDF of the unobserved responses is approximated by Gaussian distributions, whose mean and covariance are given in Algorithm 1.

As stated earlier, the EM algorithm converges to a local optimum under some normality conditions [54]. However, suitable initial estimations of the noise parameters are required for avoiding undesirable optima. For this purpose, Appendix (B) provides a detailed algorithm for obtaining the initial estimations of the noise covariance matrices using steady-state solutions of the estimators.

**Algorithm 1.**
Proposed BEM algorithm for coupled input-state-parameter-noise identification and uncertainty quantification

1: Set the initial state mean and covariance ($\hat{\boldsymbol{\mu}}_{\xi_0}$ and $\hat{\mathbf{P}}_{\xi_0}$).

2: Set the noise covariance matrices $\hat{\mathbf{Q}}^z$, $\hat{\mathbf{Q}}^p$, and $\hat{\mathbf{R}}^a$ (May apply Algorithm 2 of Appendix (B)).

3: Set the parameter noise covariance matrix, e.g., $\hat{\mathbf{Q}}^\theta = 10^{-8}\mathbf{I}_{N_\theta}$, and construct $\hat{\mathbf{Q}}^a = \text{block-diag}[\hat{\mathbf{Q}}^z, \hat{\mathbf{Q}}^\theta, \hat{\mathbf{Q}}^p]$

**EM algorithm:**

4: Set the convergence tolerance ($TOL$), e.g., $10^{-4}$; the maximum number of iterations ($ITRMAX$), e.g., 200.

5: Set the iteration number ($j$), the convergence metric ($CON$), and the surrogate function ($\hat{L}_0$) to 1.

6: **While** ($CON < TOL$) **or** ($j < ITRMAX$) {

7:     **For** $k=1:n$ {**E-Step:** Run the EKF along with the fixed-point smoother based on Eqs. (39-45):

8:         $EKF$ : $\begin{cases} \mathbf{F}^\xi_{k-1|k-1} = \left.\dfrac{\partial f(\boldsymbol{\xi}_{k-1})}{\partial \boldsymbol{\xi}^T_{k-1}}\right|_{\boldsymbol{\xi}_{k-1}=\boldsymbol{\xi}_{k-1|k-1}} \\ \boldsymbol{\xi}_{k|k-1} = f(\boldsymbol{\xi}_{k-1|k-1}) \\ \mathbf{P}^\xi_{k|k-1} = \mathbf{F}^\xi_{k-1|k-1}\mathbf{P}^\xi_{k-1|k-1}(\mathbf{F}^\xi_{k-1|k-1})^T + \hat{\mathbf{Q}}^a \\ \mathbf{H}^\xi_{k|k-1} = \left.\dfrac{\partial h(\boldsymbol{\xi}_k)}{\partial \boldsymbol{\xi}^T_k}\right|_{\boldsymbol{\xi}_k=\boldsymbol{\xi}_{k|k-1}} \\ \mathbf{G}^\xi_k = \mathbf{P}^\xi_{k|k-1}(\mathbf{H}^\xi_{k|k-1})^T(\hat{\mathbf{R}}^a + \mathbf{H}^\xi_{k|k-1}\mathbf{P}^\xi_{k|k-1}(\mathbf{H}^\xi_{k|k-1})^T)^{-1} \\ \boldsymbol{\xi}_{k|k} = \boldsymbol{\xi}_{k|k-1} + \mathbf{G}^\xi_k(\mathbf{d}^a_k - h(\boldsymbol{\xi}_{k|k-1})) \\ \mathbf{P}^\xi_{k|k} = \mathbf{P}^\xi_{k|k-1} - \mathbf{G}^\xi_k\mathbf{H}^\xi_{k|k-1}\mathbf{P}^\xi_{k|k-1} \end{cases}$

9:         $EKS$ : $\begin{cases} \mathbf{L}^\xi_{k-1} = \mathbf{P}^\xi_{k-1|k-1}(\mathbf{F}^\xi_{k-1|k-1})^T(\mathbf{P}^\xi_{k|k-1})^{-1} \\ \boldsymbol{\xi}_{k-1|k} = \boldsymbol{\xi}_{k-1|k-1} + \mathbf{L}^\xi_{k-1}(\boldsymbol{\xi}_{k|k} - \boldsymbol{\xi}_{k|k-1}) \\ \mathbf{P}^\xi_{k-1|k} = \mathbf{P}^\xi_{k-1|k-1} + \mathbf{L}^\xi_{k-1}(\mathbf{P}^\xi_{k|k} - \mathbf{P}^\xi_{k|k-1})(\mathbf{L}^\xi_{k-1})^T \end{cases}$

10:     } **End For**

11: **M-Step:** Update $\Phi$ based on Eqs. (52-55) and provide $\hat{\Phi}^{(j)}$:

12: $\hat{\boldsymbol{\mu}}_{\xi_0} = \boldsymbol{\xi}_{0|1}$

13: $\hat{\mathbf{P}}_{\xi_0} = \mathbf{P}^\xi_{0|1}$

14: $\hat{\mathbf{R}}^a = \dfrac{1}{n}\sum_{k=1}^{n}\left[(\mathbf{d}^a_k - h(\boldsymbol{\xi}_{k|k+1}))(\mathbf{d}^a_k - h(\boldsymbol{\xi}_{k|k+1}))^T + \mathbf{H}^\xi_{k|k+1}\mathbf{P}^\xi_{k|k+1}(\mathbf{H}^\xi_{k|k+1})^T\right]$

15: $\hat{\mathbf{Q}}^a = \dfrac{1}{n}\sum_{k=1}^{n}\left[(\boldsymbol{\xi}_{k|k+1} - f(\boldsymbol{\xi}_{k-1|k}))(\boldsymbol{\xi}_{k|k+1} - f(\boldsymbol{\xi}_{k-1|k}))^T + \mathbf{P}^\xi_{k|k+1} + \mathbf{F}^\xi_{k-1|k-1}\mathbf{P}^\xi_{k-1|k}(\mathbf{F}^\xi_{k-1|k-1})^T - \mathbf{F}^\xi_{k-1|k-1}\mathbf{L}^\xi_{k-1}\mathbf{P}^\xi_{k-1|k} - \mathbf{P}^\xi_{k-1|k}(\mathbf{L}^\xi_{k-1})^T(\mathbf{F}^\xi_{k-1|k-1})^T\right]$

16: Calculate $\hat{L}_1 = L(\hat{\Phi}^{(j)} | \hat{\Phi}^{(j-1)})$ by replacing $\hat{\Phi}^{(j)}$ into Eq. (37).

17: Calculate $CON = (\hat{L}_1 - \hat{L}_0)/\hat{L}_0$.

18: Set $\hat{L}_0 = \hat{L}_1$ and $j = j+1$.

19: } **End While**

**Virtual sensing of dynamical responses:**

20: **For** $k=1:n$ {Compute the posterior distribution: $p(\mathbf{d}^e_k | D_n) = N(\mathbf{d}^e_k | \mathbf{d}^e_{k|n}, \mathbf{P}^e_{k|n})$

21: $\mathbf{d}^e_{k|n} = \mathbf{G}_e(\boldsymbol{\theta}_{k|k+1})\boldsymbol{\xi}_{k|k+1}$

$\mathbf{P}^e_{k|n} = \left(\left.\dfrac{\partial(\mathbf{G}_e(\boldsymbol{\theta}_k)\boldsymbol{\xi}_{k|k+1})}{\partial \boldsymbol{\xi}_k}\right|_{\boldsymbol{\xi}_k=\boldsymbol{\xi}_{k|k+1}}\right)\mathbf{P}^\zeta_{k|k+1}\left(\left.\dfrac{\partial(\mathbf{G}_e(\boldsymbol{\theta}_k)\boldsymbol{\xi}_{k|k+1})}{\partial \boldsymbol{\xi}_k}\right|_{\boldsymbol{\xi}_k=\boldsymbol{\xi}_{k|k+1}}\right)^T$

22: } **End For**

## 6. Numerical Example with Synthetic Data

A classically damped 8-DOF system, displayed in Fig. 2, is chosen to demonstrate the proposed methodology. All masses, springs, and dampers of the system are identical, considered as $m_i = 1$ kg, $k_i = 1000$ N/m, and $c_i = 1$ N.s/m, where $i \in \{1, 2, ..., 8\}$. Based on these specifications, the modal frequencies of the structure would range from 0.93Hz to 9.89Hz. The mass matrix is assumed to be known, but the stiffness and damping matrices are entirely unknown. Thus, the structural parameter vector consists of 16 unknown parameters, expressed as $\boldsymbol{\theta} = [k_1, ..., k_8, c_1, ..., c_8]^T$. Two sensor setups are considered, wherein the following quantities are measured:
- Configuration (a): Acceleration responses of the 1st, 4th, and 8th DOFs, as well as displacement responses of the 1st and 4th DOFs
- Configuration (b): Acceleration responses of the 1st, 4th, and 8th DOFs

In configuration (b), input pseudo-observations are considered as specified in Section 2.2, aiming at alleviating the drift problem. Given these measurement settings, two input scenarios are considered:
- Case I: An unknown input force acting on the 1st DOF
- Case II: Two unknown input forces acting on the 1st and 4th DOFs

In both input scenarios, no external forces are considered on other DOFs. Both sensor placements well collocate with the external input forces, which is an effective strategy to achieve better accuracy. In this paper, because the location of unknown forces was given, collocated sensor configurations are naturally preferred as the objective is to showcase the performance of the BEM. Although using a non-collocated sensor placement does not lead to any theoretical instability in the proposed BEM approach, the accuracy of input estimations might be affected to some extent [8,14].

The synthetic response of this system is generated considering 0.001s intervals. The measurement noise is modeled by a zero-mean GWN with a standard deviation equal to 1% root-mean-square (RMS) of the noise-free responses.

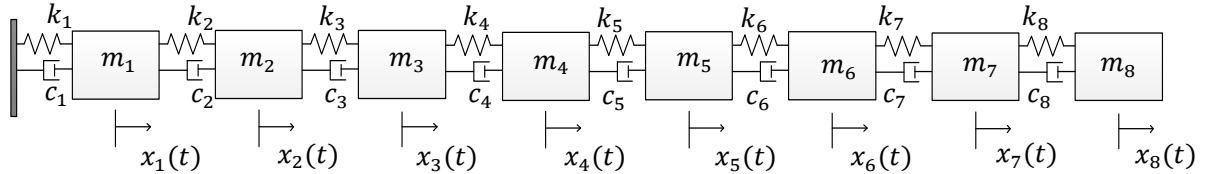

**Fig. 2.** The linear structure considered to investigate the accuracy of the proposed method

Having considered these settings, we aim to apply the BEM for estimating dynamical responses, structural parameters (stiffness and damping), and input forces. Before moving to the results, the ORC test is performed to assure theoretical observability of these quantities for both sensor setups. Fig. 3 shows the rank of the observability matrix considering the foregoing setups. When two displacement responses are considered in addition to the three accelerations, the observability improves from 15th-row to 11th-row for the input scenario (I), as shown in Fig. 3(a). Additionally, the observability improves from unobservable to 15th-row observable for the input scenario (II), as illustrated in Fig. 3(b). When the observability of sensor configuration (b) is tested along with the input pseudo-data, the system turns out to be 10th-row observable in both input scenarios. In the ORC test, the pseudo-observations play out similar to real measurements, so the rank conditions correspond to the best achievable performance that one would expect from the incorporation of pseudo-observations.

The observability results, shown in Fig. 3, can also be interpreted in terms of the effects of unknown input forces. When displacement measurements are available (Configuration a), and only one unknown input force exists, the system is 11th-row observable. However, an additional input force significantly weakens the observability, shifting to 15th-row observable. These effects are more pronounced when only acceleration measurements are available (Configuration b). For sensor Configuration (b), an additional unknown input force considerably reduces the observability.

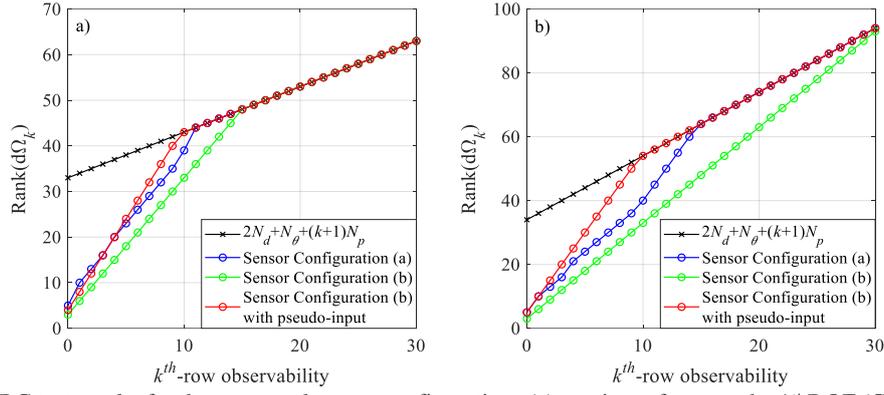

**Fig. 3.** The ORC test results for the presumed sensor configurations (a) one input force on the 1st DOF (Case I) (b) Two input forces on the 1st and 4th DOFs (Case II)

Having analyzed the observability, simulation results are presented for the selected cases of input scenarios and sensor configurations. To run Algorithms 1 and 2, the following assumptions were made:
- The stiffness and damping parameters are set to 900 N/m and 1.1 N.s/m at the initialization stage
- The structure is considered to be initially at rest.
- The noise covariance matrices are initially set at $\mathbf{Q}^z = 10^{-13}\mathbf{I}_{16}$, $\mathbf{Q}^p = 10^3$, and $\mathbf{Q}^\theta = 10^{-7}\mathbf{I}_{16}$.
- The convergence tolerance is set at $2\times 10^{-4}$.

### 6.1. Case I: GWN Input Force
*6.1.1. Main Results and Discussions*
A GWN input force with zero mean and 5N standard deviation is assumed to act on the 1st DOF. We use noisy response measurements of the system to estimate the states, inputs, structural parameters, and noise characteristics. Algorithm 2 (see Appendix B) is applied for calculating the initial estimates of the noise parameters, and then, Algorithm 1 is used for solving the coupled estimation problem. Both algorithms have converged quickly within a few iterations.

All eight springs' stiffness is identified using the noisy data obtained from both setups, and the results are shown in Fig. 4. The estimated mean of the stiffness parameters is in close agreement with the actual values for both sensor configurations. While the results of the two sensor setups slightly differ for $k_1$, both lead to the same mean and uncertainty bound for $k_2$ to $k_8$. The uncertainty associated with $k_1$ appears to be larger than other parameters, which is attributed to having an unknown input force on the 1st DOF. This observation makes intuitive sense as the presence of unknown input induces additional uncertainties.

Estimations of the damping coefficients are shown in Fig. 5. Both sensor setups lead to accurate and similar estimations in terms of the mean and uncertainty bounds for $\{c_2, c_3, ..., c_8\}$. However, the accuracy of $c_1$ is not as good as other damping coefficients, mainly because the unknown force was applied to the 1st DOF. This result reveals the weaker observability of $c_1$ and $k_1$ compared to all other unknown parameters.

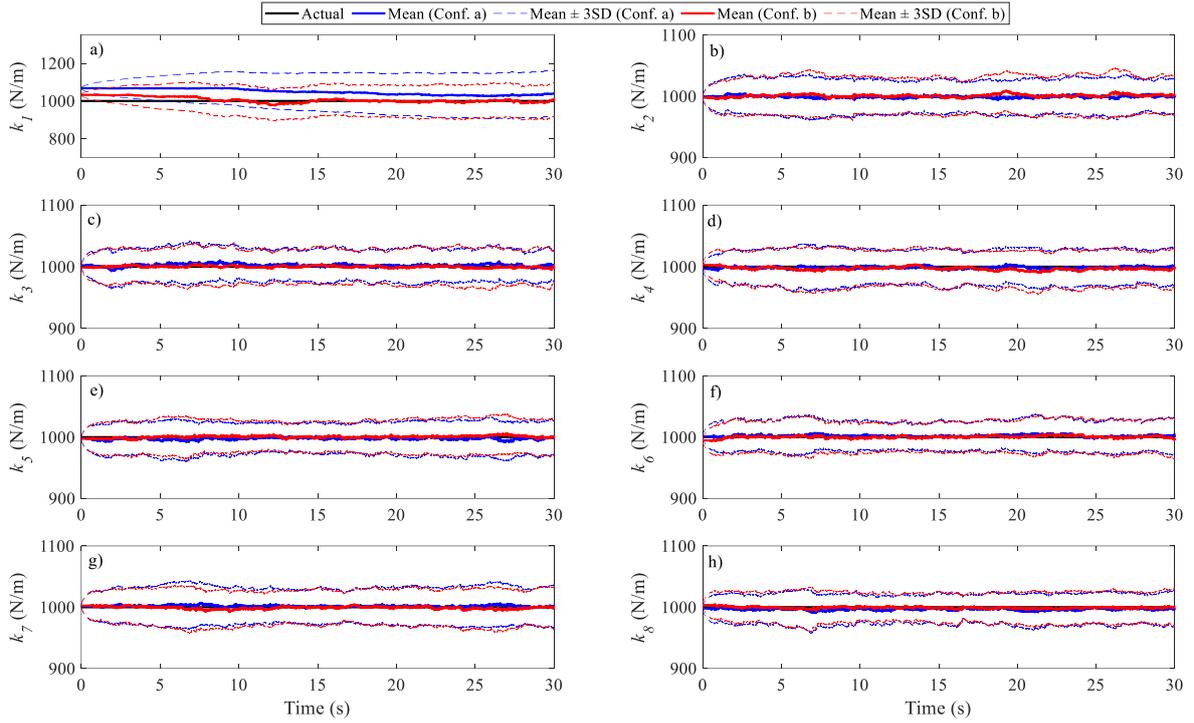

**Fig. 4.** Identification of the springs' stiffness using the foregoing sensor setups (the vertical axis limits in subfigure (a) is increased slightly to display the entire uncertainty bounds; nominal values: 1000 N/m)

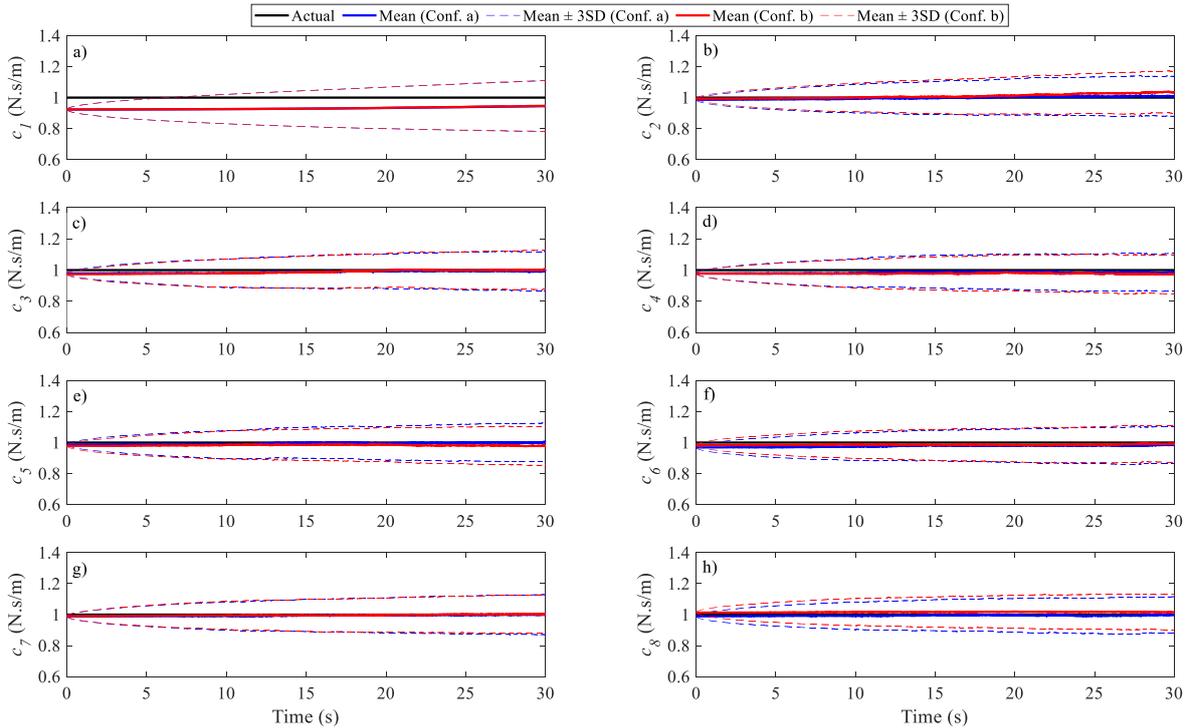

**Fig. 5.** Identification of damping coefficients using the foregoing setups (nominal values: 1 N.s/m)

Fig. 6(a-c) shows the estimations of the unknown GWN input force. The accuracy of estimations is remarkable for both setups. Fig. 6(b) displays the estimations in the frequency-domain. Comparing the results of the two configurations, it is clear that the low-frequency components are better matched when displacement measurements are considered. In configuration (a), the estimated mean matches better with the actual response, offering reasonably small uncertainty bounds. This conclusion is reinforced when comparing the estimation errors in Fig. 6(c), where the inclusion of displacement

responses in configuration (a) leads to smaller errors. For both setups, it is important to note that the estimation errors entirely fall within the corresponding uncertainty bounds.

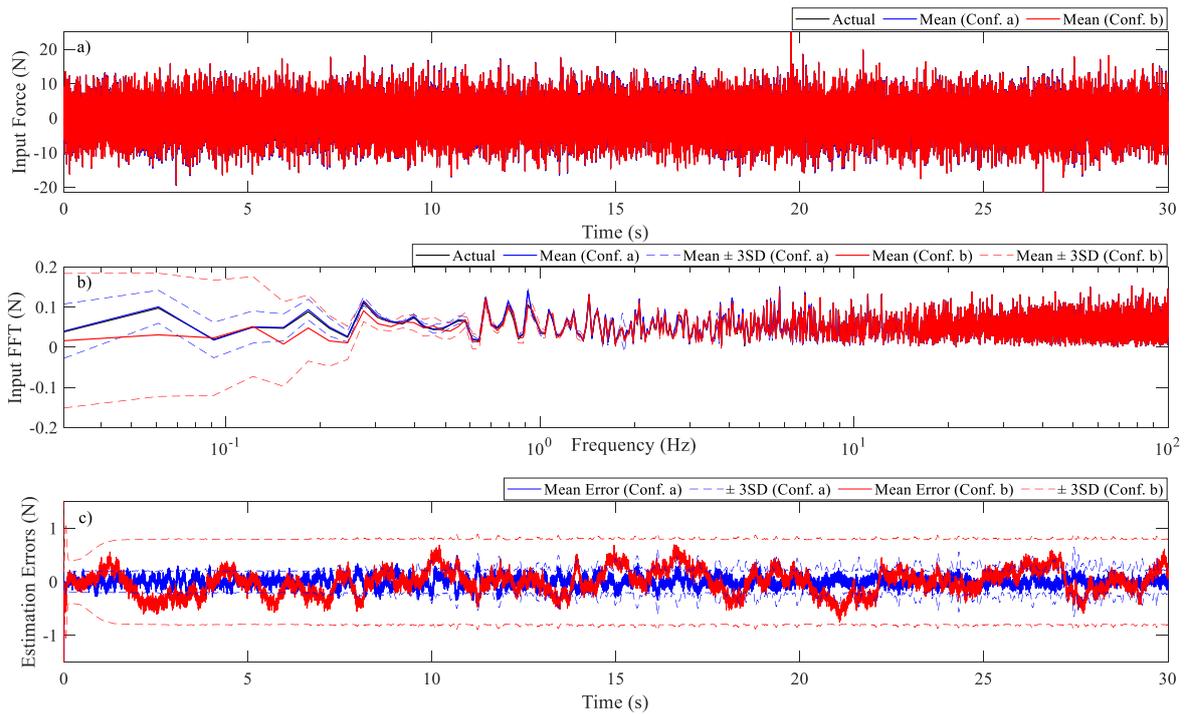

**Fig. 6.** Identification of GWN input force using the presumed sensor configurations (a) estimations in time-domain (b) estimations in frequency-domain (c) estimation errors

In Fig. 7, the dynamical responses of the 6$^{th}$ DOF are plotted, which showcases the accuracy in this unobserved DOF. Both configurations provide almost identical mean and uncertainty bounds. However, setup (a) outperforms (b) in predicting the displacement response of the 6$^{th}$ DOF. This observation was expected as setup (a) includes displacement measurements as well, which improves the identification of low-frequency components. Another notable observation is that the uncertainty bounds of each setup well account for the corresponding estimation errors.

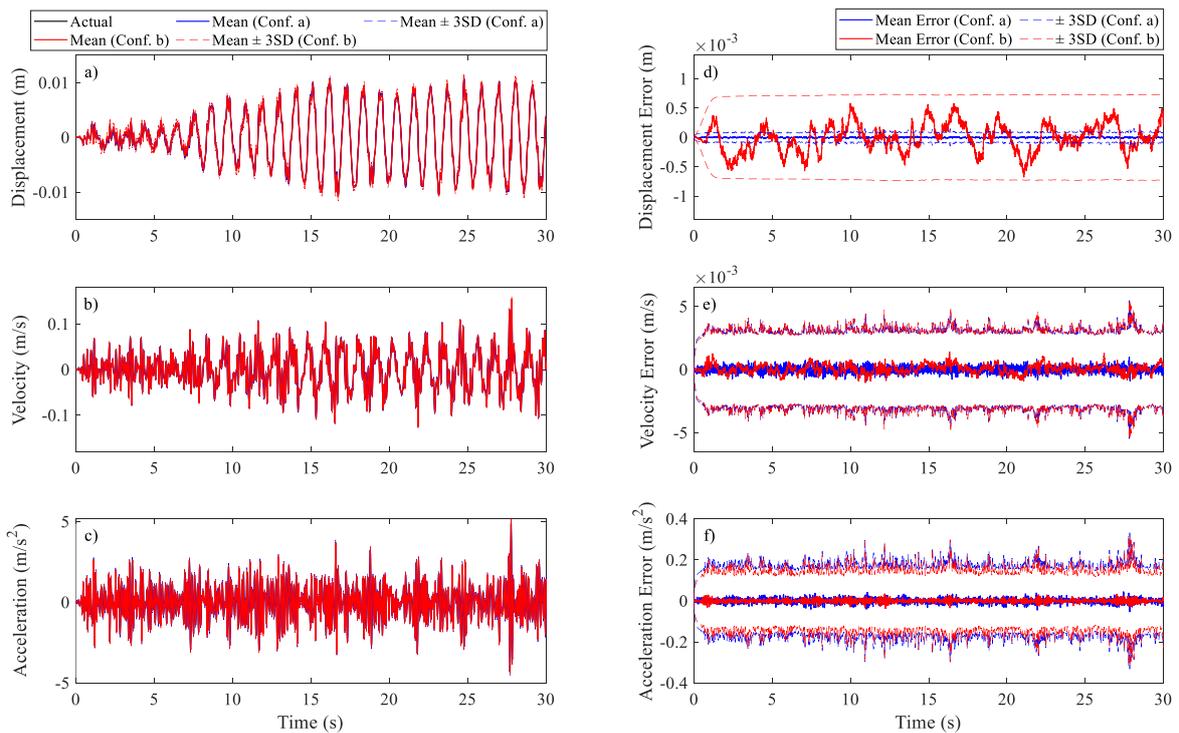

**Fig. 7.** Response predictions of the 6th DOF (a-c) displacement, velocity, and acceleration histories (d-f) estimation errors of the same quantities

Fig. 8(a) shows that the substitute function $L(\Lambda|\hat{\Lambda}^{(j-1)})$ stabilizes and converges after 4 iterations of Algorithm 2 (see Appendix B). This algorithm also provides reasonable starting values for the noise parameters, as shown in Fig. 8(b-c). When Algorithm 1 is applied, the function $L(\Phi|\hat{\Phi}^{(j-1)})$ increases considerably in the first iteration and then stabilizes, as shown in Fig. 8(d). All components of the measurement noise covariance matrix converge to the nominal values, except for the noise variance of the 1st DOF displacement response ($R^a_{(1,1)}$), as indicated in Fig. 8(e). This component of the measurement noise covariance converges to a value larger than its actual one, which is attributed to the contribution of input uncertainty to the response of this DOF. Some components of the process noise covariance matrix are shown in Fig. 8(f). Specifically, the input noise variance ($Q^p_{(1,1)}$) converges to 49.44. Moreover, the state noise variances ($Q^z_{(1,1)}, Q^z_{(8,8)}, Q^z_{(9,9)}$ and $Q^z_{(16,16)}$) stabilize and reach non-trivial values, which is hard to achieve through most existing methods. However, since there is no reference value to compare the results with, it is difficult to confirm whether the estimations are correct. Nevertheless, the convergence property of the results is definitely remarkable in this case.

Fig. 9 shows the results for sensor configuration (b) when the dummy-input is also considered, which further confirms the preceding conclusions about the identification of the process noise parameters. Additionally, the noise variance of the dummy-input is estimated to be 24.90 N, which is close to the actual variance of the GWN input force (25 N). It is also observed that the steady-state initializer well estimates the measurement noise covariance matrices, and the main BEM algorithm improves them considerably.

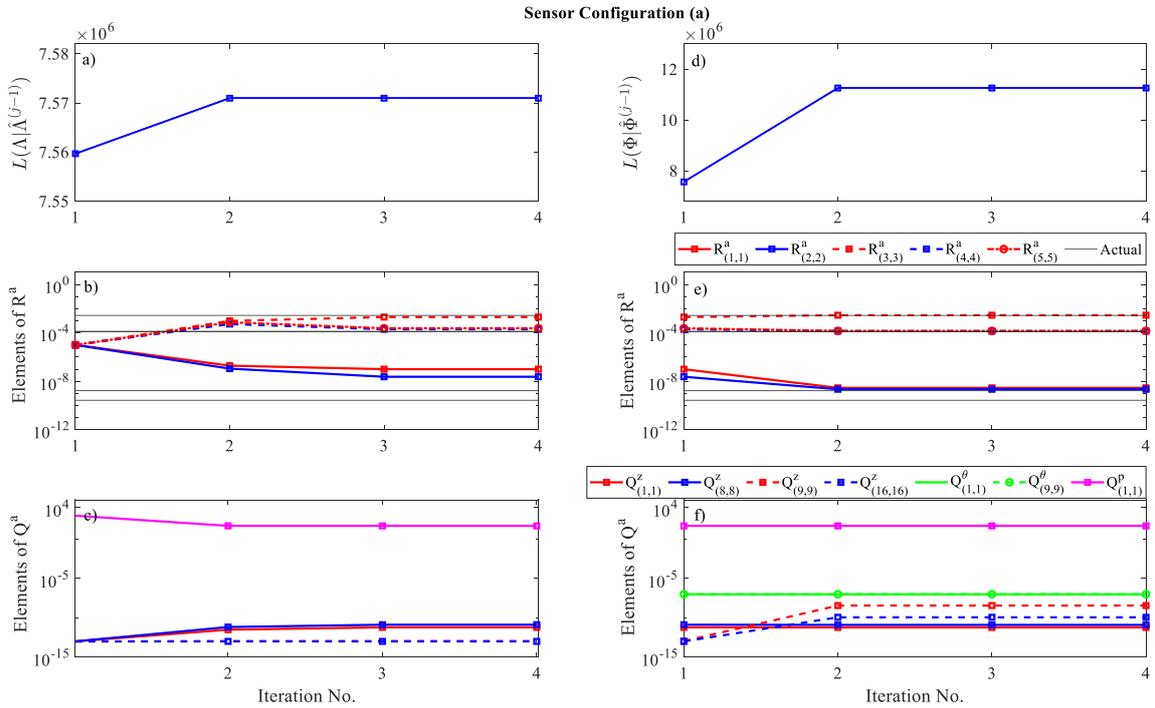

**Fig. 8.** Variation of the substitute function and the noise parameters when using data from sensor setup (a) ($R^a_{(1,1)}$ and $R^a_{(2,2)}$ indicate the measurement noise variances of the 1st and 4th DOFs' displacement responses, respectively; $R^a_{(3,3)}$, $R^a_{(4,4)}$, and $R^a_{(5,5)}$ represent the measurement noise variances of the 1st, 4th, and 8th DOFs' acceleration responses, respectively)

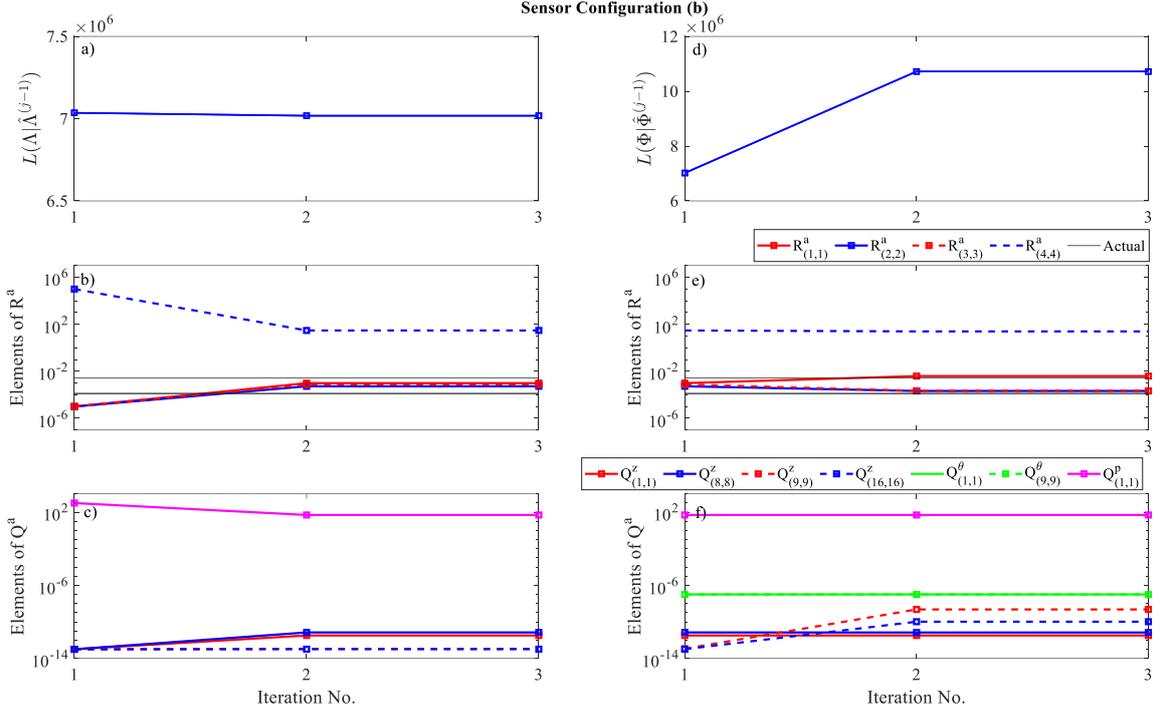

**Fig. 9.** Variation of the substitute function and the noise parameters when using data from sensor setup (b) ( $R^a_{(1,1)}$, $R^a_{(2,2)}$, and $R^a_{(3,3)}$ represent the measurement noise variances of the 1st, 4th, and 8th DOFs' acceleration responses, respectively; $R^a_{(4,4)}$ is the dummy-input noise variance)

*6.1.2. Sensitivity to initial choices of covariance matrices*

Although the BEM optimizes the noise parameters, it requires a reasonable starting point similar to most optimization tools. In Table 1, seven combinations of initial covariance matrices are considered for sensor configurations (a) and (b). Then, the BEM algorithm is used to identify the unknown quantities. The estimated noise parameters are plotted in Fig. 10(a-b) for different starting points. It is evident that the displayed noise parameters reach the same neighborhood regardless of the starting points. However, there exists slight variation in some of the identified parameters. Such variation can stem from the difference in the trajectory of searching algorithm, and the fact that the convergence metric might terminate the algorithm at a slightly different point. Perhaps, a tighter stopping criterion can reduce this deviation in the results, but it comes at the expense of additional iterations, causing a rise in the computational cost. However, in this section, we did not change the above convergence metric to retain consistency in reporting the results.

**Table 1.** Different choices of covariance matrices fed into Algorithms 1 and 2 for investigating starting point effects

| Noise Parameters | Combination No. | | | | | | |
| --- | --- | --- | --- | --- | --- | --- | --- |
| | 1 | 2 | 3 | 4 | 5 | 6 | 7 |
| | Sensor Configuration (a) | | | | | | |
| $\mathbf{R}$ | $10^{-5} \times \mathbf{I}_5$ | $10^{-3} \times \mathbf{I}_5$ | $10^{-2} \times \mathbf{I}_5$ | $10^{-5} \times \mathbf{I}_5$ | $10^{-5} \times \mathbf{I}_5$ | $10^{-5} \times \mathbf{I}_5$ | $10^{-7} \times \mathbf{I}_5$ |
| $\mathbf{Q}^z$ | $10^{-13} \times \mathbf{I}_{16}$ | $10^{-10} \times \mathbf{I}_{16}$ | $10^{-10} \times \mathbf{I}_{16}$ | $10^{-12} \times \mathbf{I}_{16}$ | $10^{-12} \times \mathbf{I}_{16}$ | $10^{-14} \times \mathbf{I}_{16}$ | $10^{-14} \times \mathbf{I}_{16}$ |
| $\mathbf{Q}^p$ | $10^3$ | $10^3$ | $10^1$ | $10^3$ | $10^{-1}$ | $10^3$ | $10^2$ |
| | Sensor Configuration (b) | | | | | | |
| $\mathbf{R}$ | $10^{-5} \times \mathbf{I}_3$ | $10^{-3} \times \mathbf{I}_3$ | $10^{-2} \times \mathbf{I}_3$ | $10^{-5} \times \mathbf{I}_3$ | $10^{-5} \times \mathbf{I}_3$ | $10^{-5} \times \mathbf{I}_3$ | $10^{-7} \times \mathbf{I}_3$ |
| $\mathbf{R}^{pd}$ | $10^5$ | $10^5$ | $10^5$ | $10^3$ | $10^5$ | $10^5$ | $10^5$ |
| $\mathbf{Q}^z$ | $10^{-13} \times \mathbf{I}_{16}$ | $10^{-10} \times \mathbf{I}_{16}$ | $10^{-10} \times \mathbf{I}_{16}$ | $10^{-12} \times \mathbf{I}_{16}$ | $10^{-12} \times \mathbf{I}_{16}$ | $10^{-14} \times \mathbf{I}_{16}$ | $10^{-14} \times \mathbf{I}_{16}$ |
| $\mathbf{Q}^p$ | $10^3$ | $10^3$ | $10^1$ | $10^3$ | $10^{-1}$ | $10^3$ | $10^2$ |

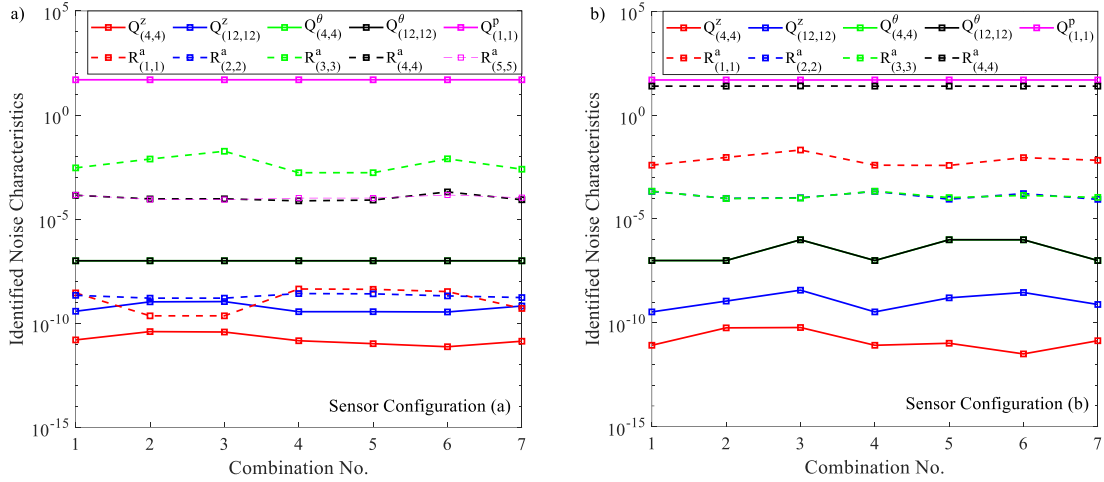

**Fig. 10.** Effects of initial choices of the noise parameters on the identified covariance matrices (notations are the same as Figs. 8 and 9)

*6.1.3. Sensitivity to initial choices of structural parameters*
From a theoretical standpoint, since observability conditions are satisfied, the unknown parameters in-limit converge to a stationary point. However, when the initial choice of the parameters changes, the trajectory of parameter estimation differs accordingly. In this case, within a limited period, the solution might look slightly different while fallen within a specific neighborhood covered by the uncertainty bounds. Therefore, different initial values might not lead to the same value within a relatively short period, but they can come close to the same neighborhood.

To contextualize this discussion, we re-run the BEM algorithm for two different initial values of $k_2$, considering $k_2^{(0)} = 800$ and $1200$ N/m. Fig. 11 shows the identified stiffness in the first and last iteration of the BEM algorithm. As indicated in Fig. 11(a), the posterior distribution represented by the mean and uncertainty bounds gradually converge to the same solution in the first iteration of the BEM. The difference almost vanishes in the next iterations of the BEM such that the results corresponding to different initial values are indistinguishable in the 5th iteration, as evidenced in Fig. 11(b). Note that such minor deviations in the estimated mean does not cause concerns in this study as long as the posterior uncertainties can cover them.

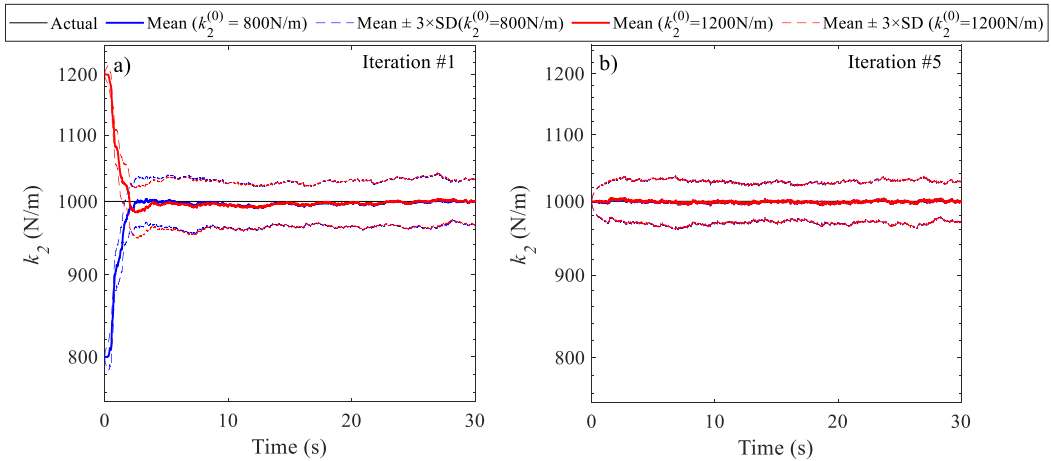

**Fig. 11.** Identification of the stiffness parameter $k_2$ using two different starting values in the BEM algorithm

### 6.2. Case I: Abrupt Stiffness Reduction
It is important to assess the performance of the BEM when an abrupt drop in the stiffness occurs. To simulate this damage scenario, at $t = 15$s, the stiffness $k_6$ is assumed to change from 1000 N/m to 750 N/m. The aforementioned settings are used for creating noisy synthetic responses of the structure under the same GWN input force. Then, the BEM algorithm is adopted to identify the input-state-parameter-noise simultaneously. Fig. 12 shows the estimations of the stiffness parameters obtained using both setups. As can be seen, the drop in the stiffness is captured well even in the absence of displacement measurements, and the disturbances caused by the damage disappear after a short period of time. Other results, including the estimations of the states, the input, and damping parameters are qualitatively good despite having non-smooth dynamics. However, they are not shown herein to avoid repetition.

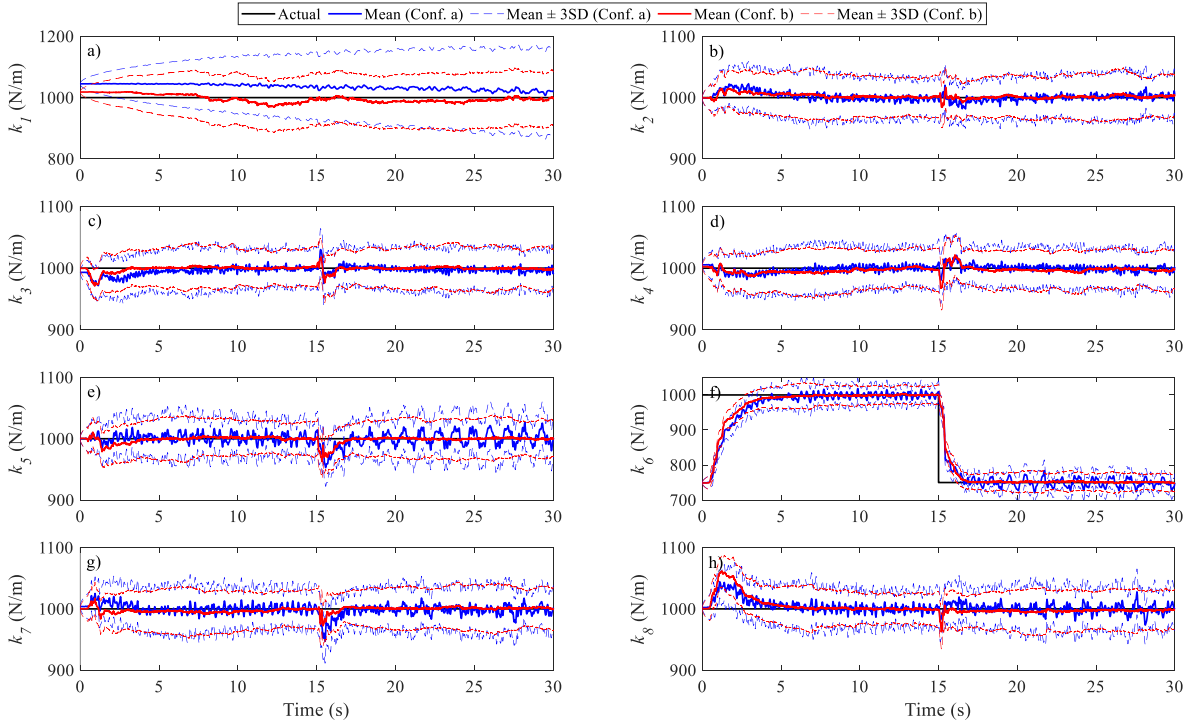

**Fig. 12.** Identification of the springs' stiffness in the presence of an abrupt reduction in $k_6$

## 6.3. Case II: Multiple Input Forces
### 6.3.1. Combination of GWN and Impact Forces
In this section, the applicability of the proposed method is investigated by assuming a combination of stationary and non-stationary forces acting on the system. This situation is simulated by assuming an unknown impact force acting on the 4th DOF at $t = 15s$ in addition to the foregoing GWN input force applied to the 1st DOF. For brevity, the results of configuration (a) are only showed since configuration (b) comparatively yields similar results as Section 6.1.

Fig. 13 depicts the identification of input forces in both time and frequency domains. As shown, the BEM is able to identify both forces accurately. The estimated mean matches well with the actual values, and the estimation errors fall within the uncertainty bounds.

Fig. 14 shows the identified springs' stiffness. The impact force creates some disturbances in $k_2$, $k_3$, $k_4$, and $k_5$, which fade out gradually. For both stiffness and damping, the BEM provides reasonable posterior PDFs, despite having short-length response measurements. The abrupt disturbances in the stiffness parameters can be discussed in terms of observability results, where the system is weakly locally observable. In the absence of strong observability, the augmented state vector cannot be identified instantaneously, and a time lag is necessary to identify the unknown quantities [17,51]. A similar pattern is observed in Fig. 15 for the damping coefficients, although instabilities are relatively smaller in this case. Note that the slower convergence of the damping parameters is rooted in the weaker sensitivity and observability of these unknowns compared to the stiffness parameters.

Additionally, Fig. 16 displays the estimations of the unmeasured responses of the 6th DOF. It is evident that the estimated mean matches with the actual response, and their discrepancy is reliably accounted for by the uncertainty bounds.

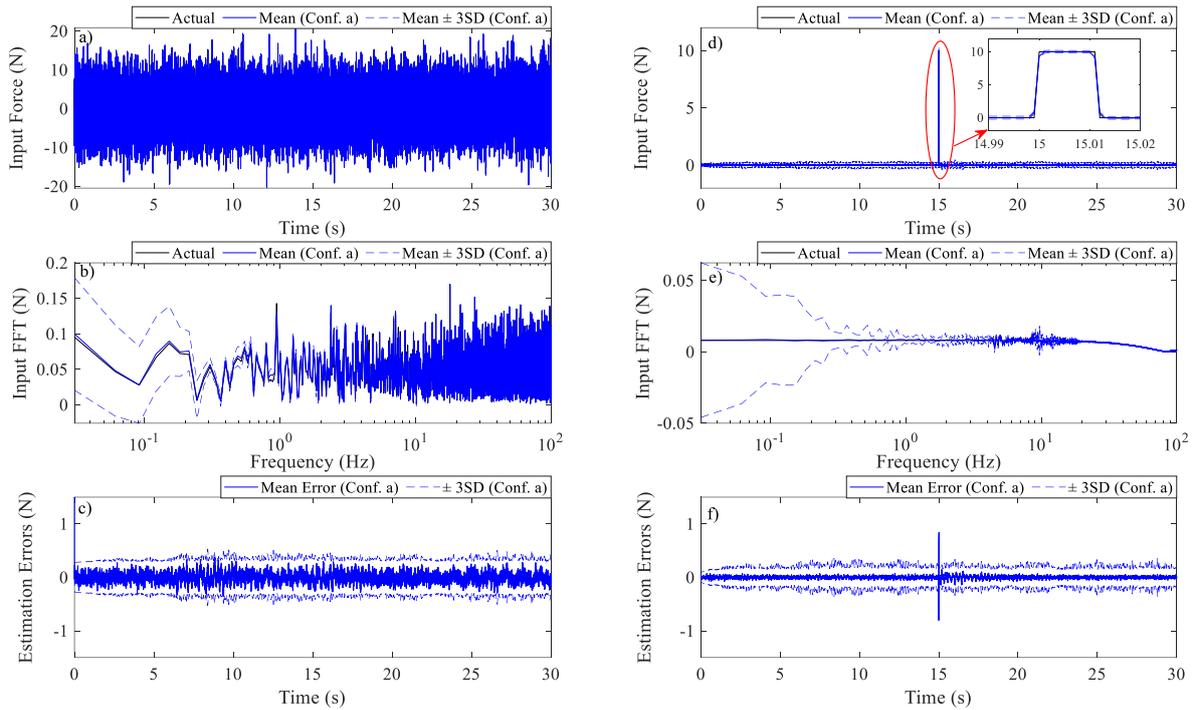

**Fig. 13.** Estimated input forces (a-c) GWN force acting on the 1st DOF (d-f) Impact force acting on the 4th DOF

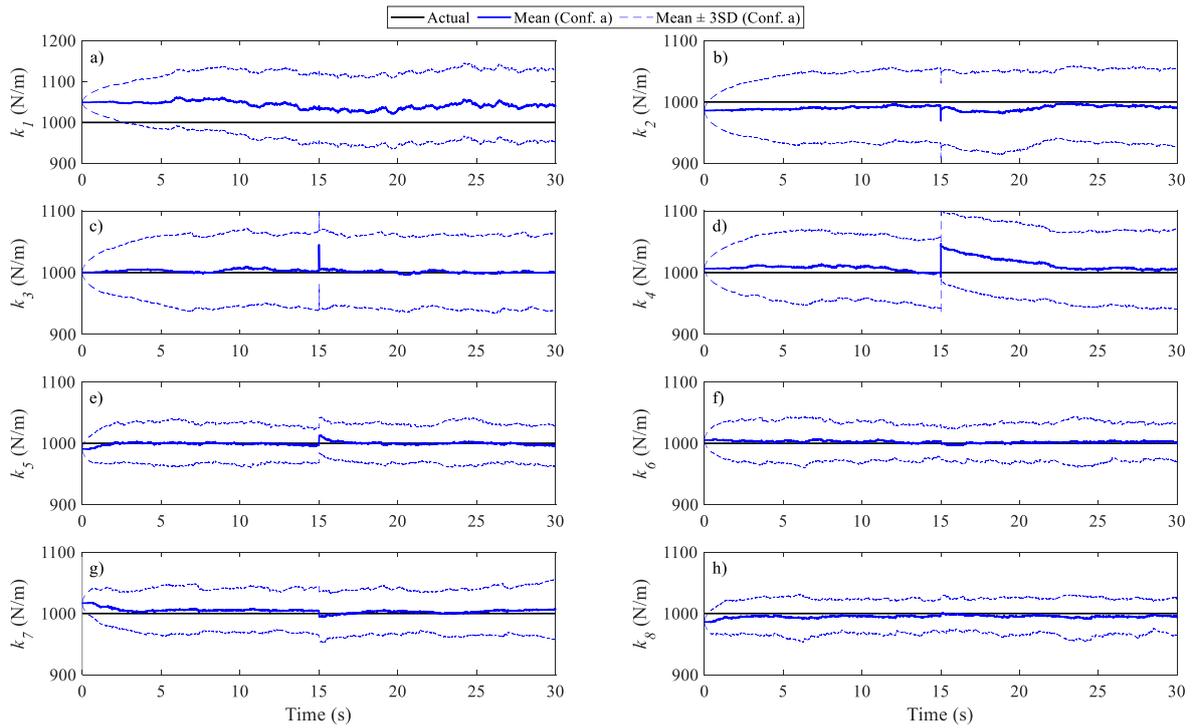

**Fig. 14.** Identification of the elements of the stiffness matrix when having two unknown forces

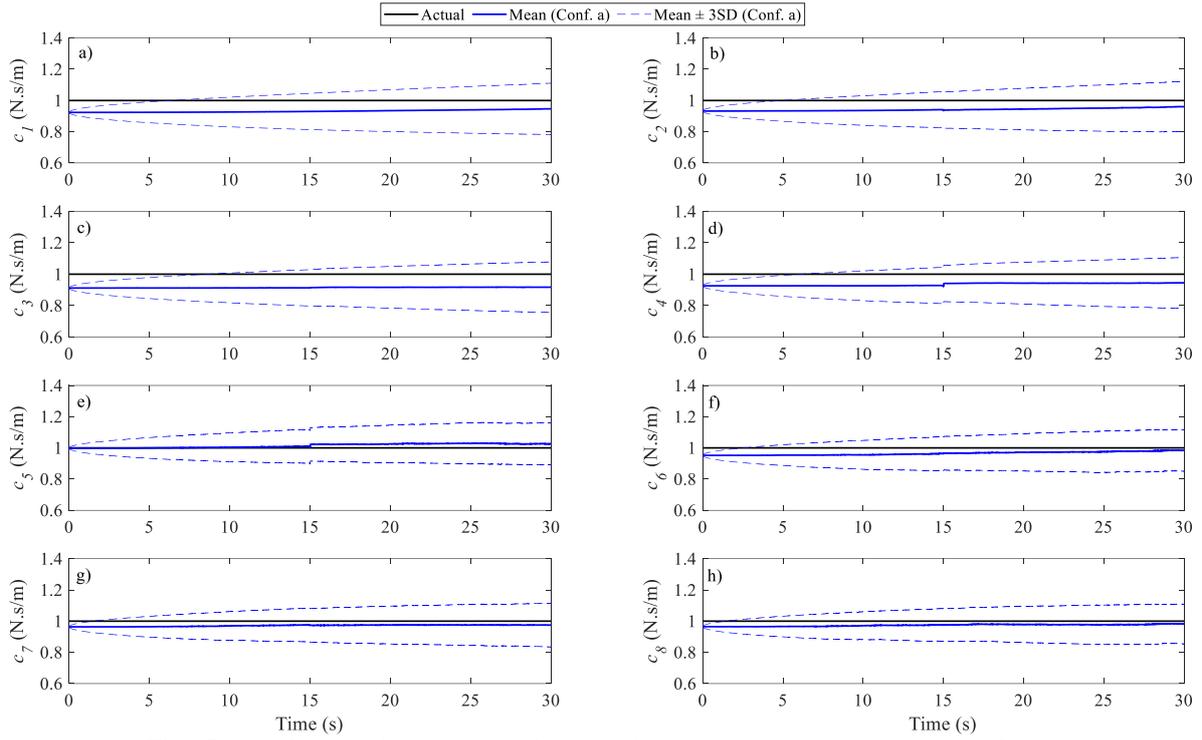

**Fig. 15.** Identification of the elements of the damping matrix considering two unknown forces

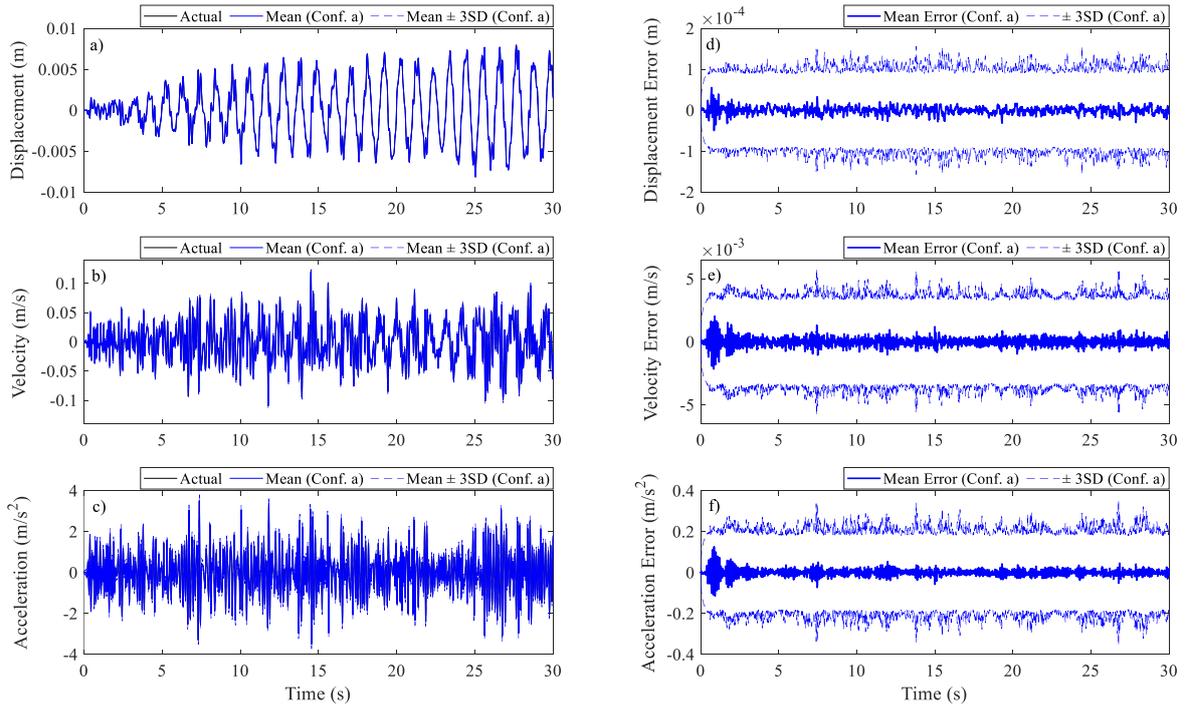

**Fig. 16.** Estimation of the unobserved responses of the 6$^{th}$ DOF

Fig. 17 shows the variation of the substitute function and the noise parameters over iterations of both Algorithms. As can be observed, the substitute function increases considerably and converges within a few iterations. The initializer provides good estimations for the measurement noise covariance matrix, which expedites the identification process by reducing the number of iterations of the main BEM algorithm. The process noise variances are also updated based on the data, reaching stable values.

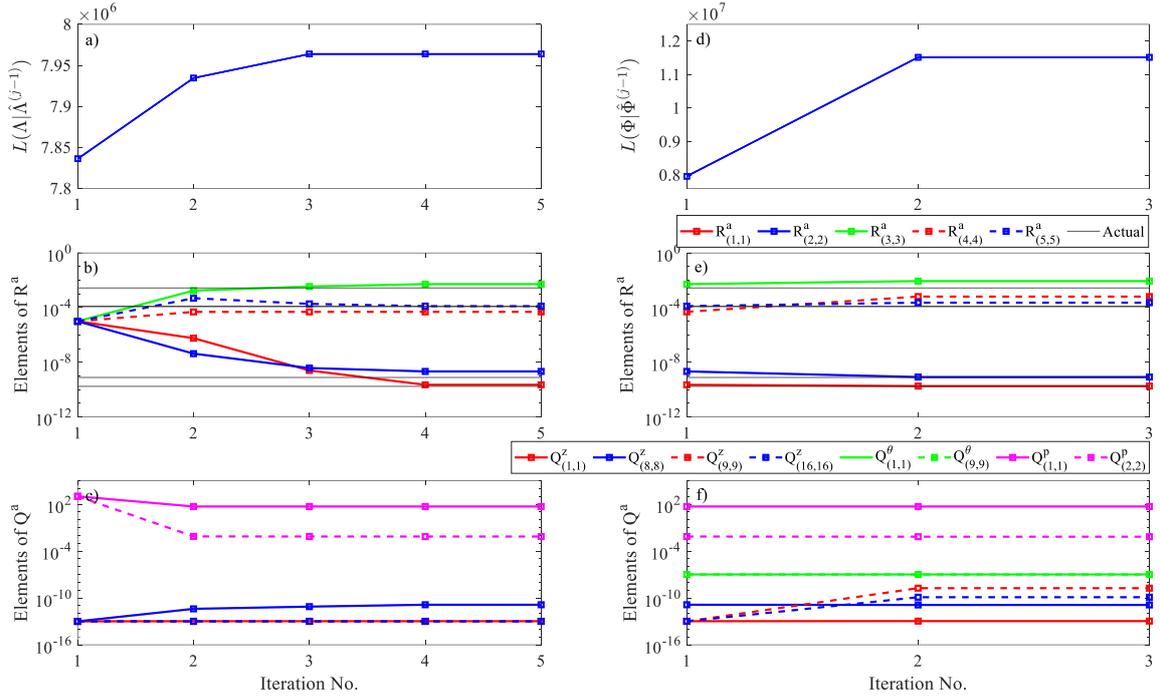

**Fig. 17.** Variation of the substitute function and the noise parameters (notations are the same as Fig. 8)

*6.3.2. Non-stationary and non-Gaussian Input Forces*

It is crucial to demonstrate identification of non-stationary and non-Gaussian input forces. For this purpose, the GWN force of the preceding section is replaced with a non-stationary narrow-band stochastic force applied to the 1st DOF. Additionally, the impact force is activated at $t = 20s$ to create a more critical non-stationary effect while still acting upon the 4th DOF. The reconstructed input forces are shown in Fig. 18 in both time and frequency domains. The input is estimated accurately, and the uncertainty bounds can account for potential discrepancies despite non-stationary and non-Gaussian effects.

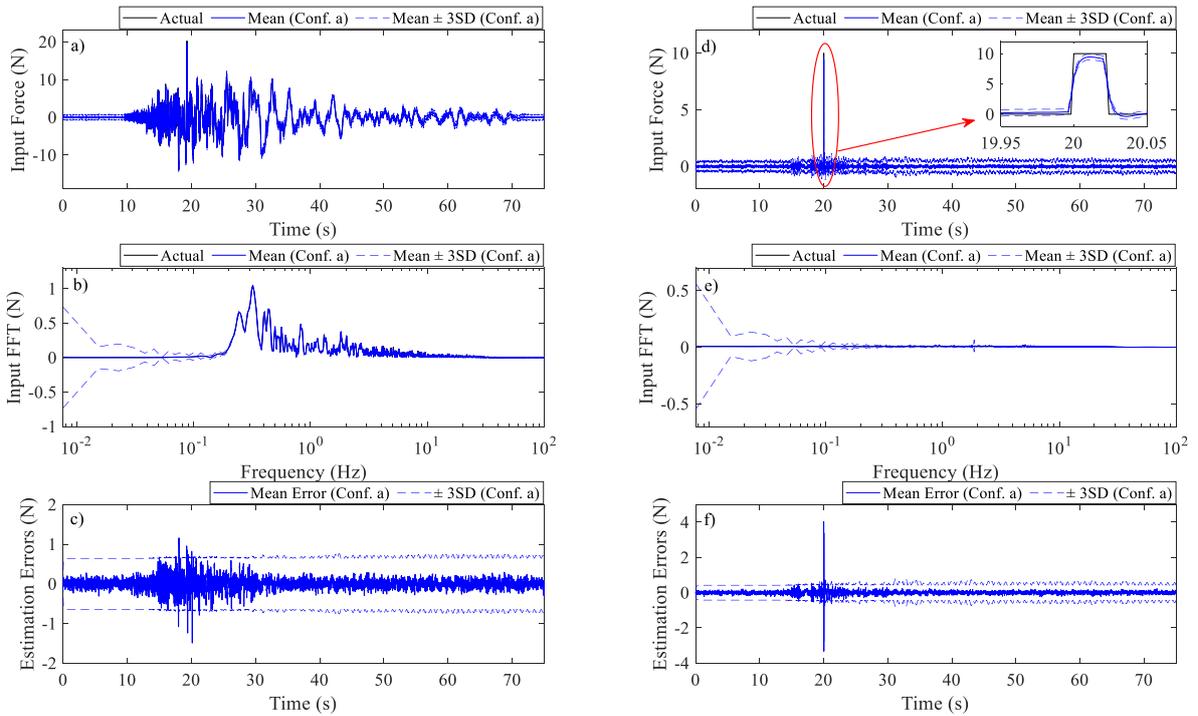

**Fig. 18.** Estimated input forces (a-c) non-stationary stochastic force acting on the 1st DOF (d-f) Impact force acting on the 4th DOF

Fig. 19 indicates the estimates of the stiffness parameters as time elapses. Although these parameters are estimated accurately, some variation appears in the estimated mean within the interval $t \in [15-25s]$ when the non-stationary forces create extreme disturbances. The estimates become more stable and smoother after a short while, and the abrupt variations vanish over time. Fig. 20 shows the identified damping parameters. In the case of damping parameters, the instabilities due to non-stationary input forces are less noticeable compared to the stiffness parameters. However, the quality of identified damping parameters is inferior compared to the stiffness parameters, which can be attributed to weaker sensitivity to the data.

Due to the loss of strong observability, it is natural to observe that the stochastic effects of input forces translate into the estimates of the unknown parameters for a short period. However, as data accumulates, the proposed method starts distinguishing the effects of unknown input forces and recovers the accuracy.

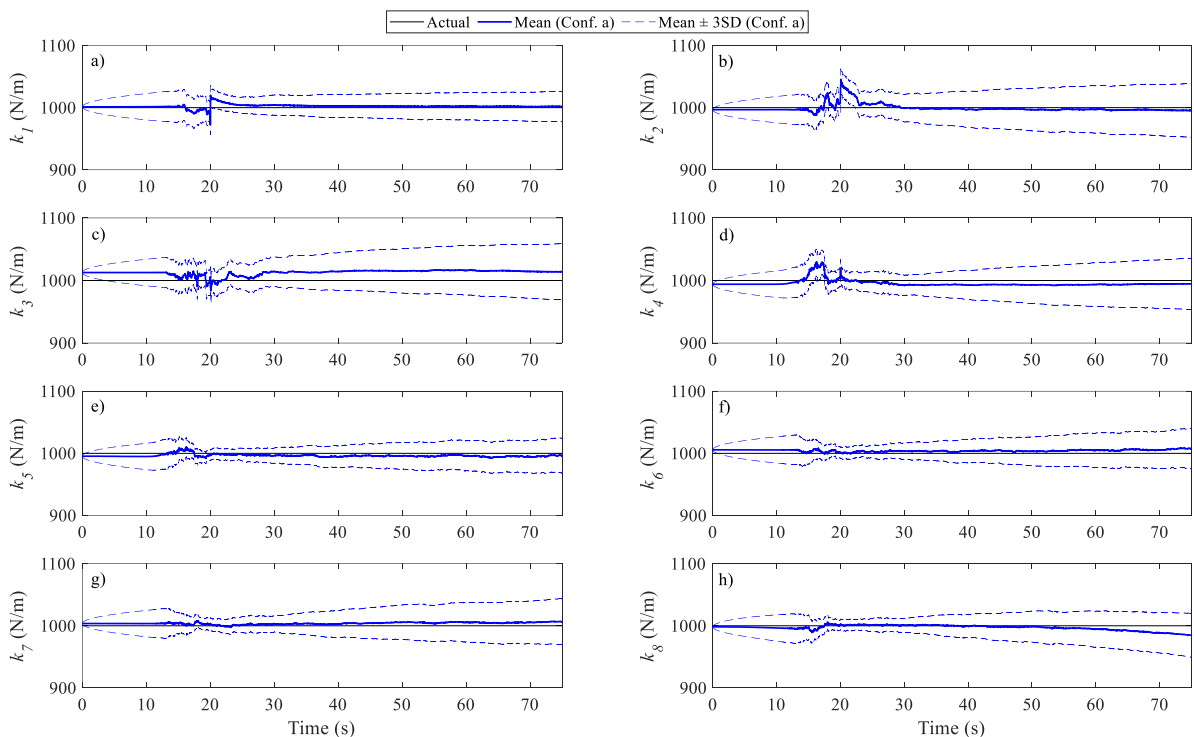

**Fig. 19.** Identification of the elements of the stiffness matrix when having two non-stationary narrow-band forces

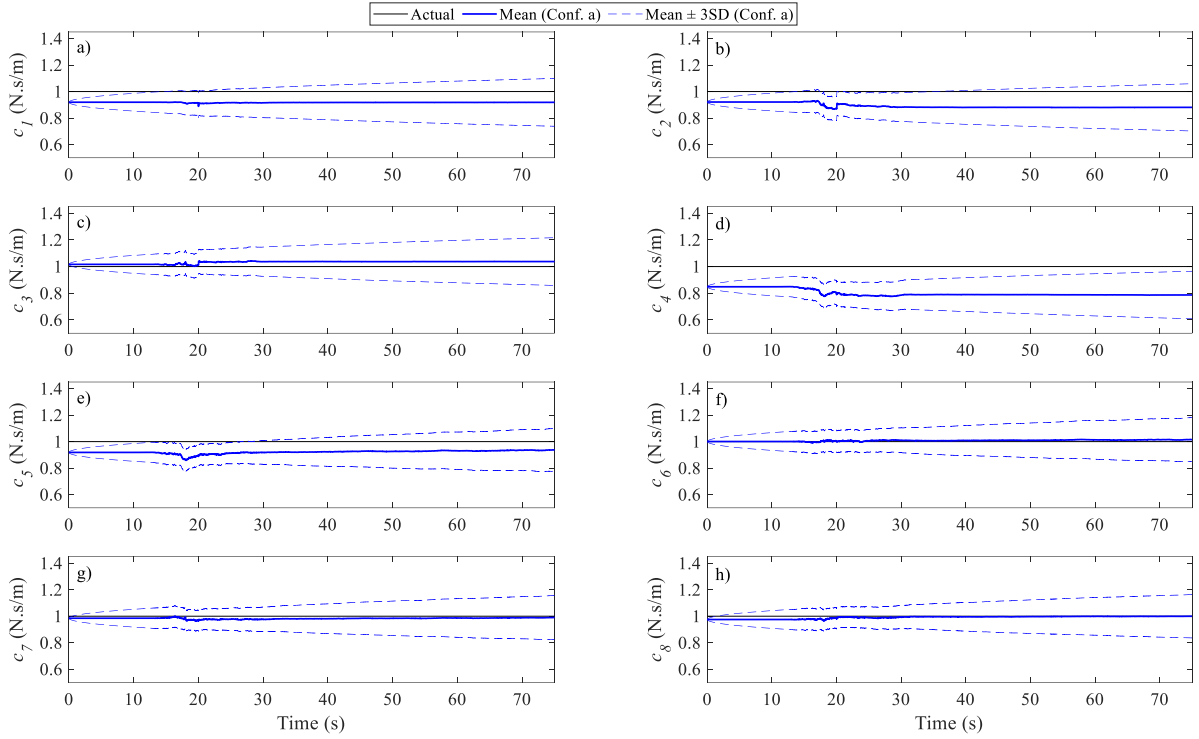

**Fig. 20.** Identification of the elements of the damping matrix when having two non-stationary narrow-band forces

## 7. Experimental Example

The small-scale frame shown in Fig. 21(a) was tested subjected to a unidirectional base excitation along the X-axis. As verified in our previous works [57], the shear frame shown in Fig. 21(b) can well describe the dynamical behavior of this structure. The mass of the structure is considered to be lumped at floor levels, given as $m_1 = 5.63 kg$, $m_2 = 6.03 kg$, $m_3 = 4.66 kg$. Further information about the structure and its dynamical model can be found elsewhere [57].

It is straightforward to characterize the stiffness matrix in terms of unknown stiffness components $\{k_1, k_2, k_3\}$. The damping of the model is considered to be proportional to the mass matrix and parameterized in terms of three modal damping ratios, i.e., $\{\xi_1, \xi_2, \xi_3\}$. This specification implies that the structural parameter vector comprises six unknowns, given as $\mathbf{\theta} = [k_1\ k_2\ k_3\ \xi_1\ \xi_2\ \xi_3]^T$. Thus, the unknown stiffness and damping matrices are constructed as

$$\mathbf{K}(\mathbf{\theta}) = \begin{bmatrix} k_1 + k_2 & -k_2 & 0 \\ -k_2 & k_2 + k_3 & -k_3 \\ 0 & -k_3 & k_3 \end{bmatrix} \tag{56}$$

$$\mathbf{C}(\mathbf{\theta}) = \sum_{i=1}^{3} \left( \frac{\mathbf{M}\hat{\mathbf{\phi}}_i \hat{\mathbf{\phi}}_i^T \mathbf{M}}{\hat{\mathbf{\phi}}_i^T \mathbf{M}\hat{\mathbf{\phi}}_i} \right) \left( 4\pi \hat{f}_i \xi_i \right) \tag{57}$$

where $\mathbf{M} = diag[m_1, m_2, m_3]$ is the known mass matrix; $\mathbf{K}(\mathbf{\theta}) \in \mathbb{R}^{3\times 3}$ is the stiffness matrix; $\mathbf{C}(\mathbf{\theta}) \in \mathbb{R}^{3\times 3}$ is the damping matrix; $\hat{f}_i$ and $\hat{\mathbf{\phi}}_i \in \mathbb{R}^3$ are the modal frequency and mode shape vector of the $i^{\text{th}}$ dynamical mode obtained based on the nominal values of the stiffness components.

It is desired to calibrate the stiffness and damping matrices, identify the base excitation, and estimate unmeasured responses using incomplete acceleration-only measurements. In this paper, the acceleration responses of the second and third floors are selected as the measured quantities for applying the BEM. The acceleration responses of the base and the first story are also measured and used for the verification. The length of measurements is 75 s, and the sampling rate is 200 Hz.

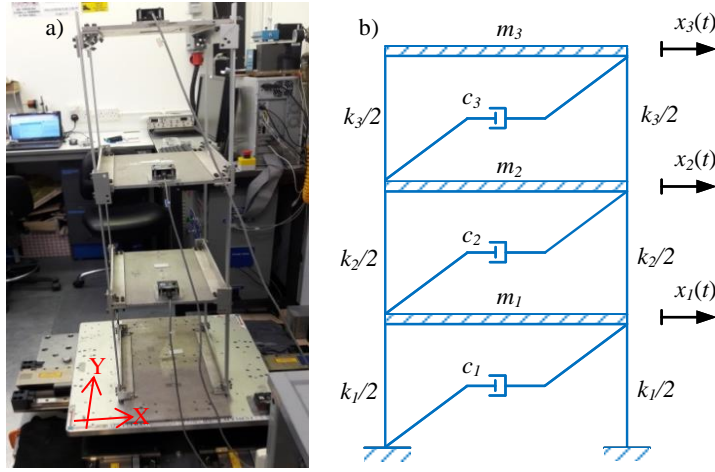

**Fig. 21.** (a) Shear frame tested under a unidirectional base excitation along the X-axis b) Structural model class

Fig. 22 shows the ORC test performed for this structure, studying input pseudo-observations. With no pseudo-data, 11$^{th}$-row observability of the system is confirmed. However, the incorporation of the pseudo-observations leads to 5$^{th}$-row observability, confirming the benefits of using the pseudo-data.

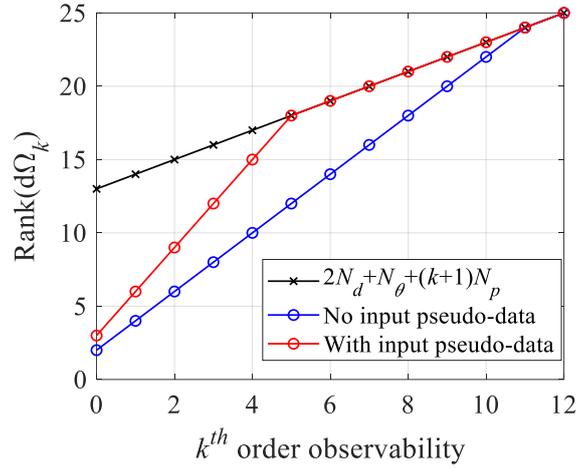

**Fig. 22.** Observability conditions for cases with and without input pseudo-data

The starting values of the noise parameters are selected as $\mathbf{Q}^z = 10^{-12}\mathbf{I}_6$, $\mathbf{Q}^{\ddot{x}_s} = 10$, $\mathbf{Q}^\theta = 10^{-7}\mathbf{I}_6$, $\mathbf{R} = 10^{-4}\mathbf{I}_2$, and $\mathbf{R}^{pd} = 10^2$. Then, the proposed algorithms are applied for the input-state-parameter identification while the noise parameters are updated iteratively. Fig. 23 shows the base excitation identified using the proposed method. Fig. 24 plots the displacement response of the 1$^{st}$ DOF to illustrate an unobserved response. Compared to the reference values, the estimations are accurate, and the uncertainty bounds well account for the discrepancies. Similar results are obtained for other elements of the state vector, but they are not provided in view of the length of the paper.

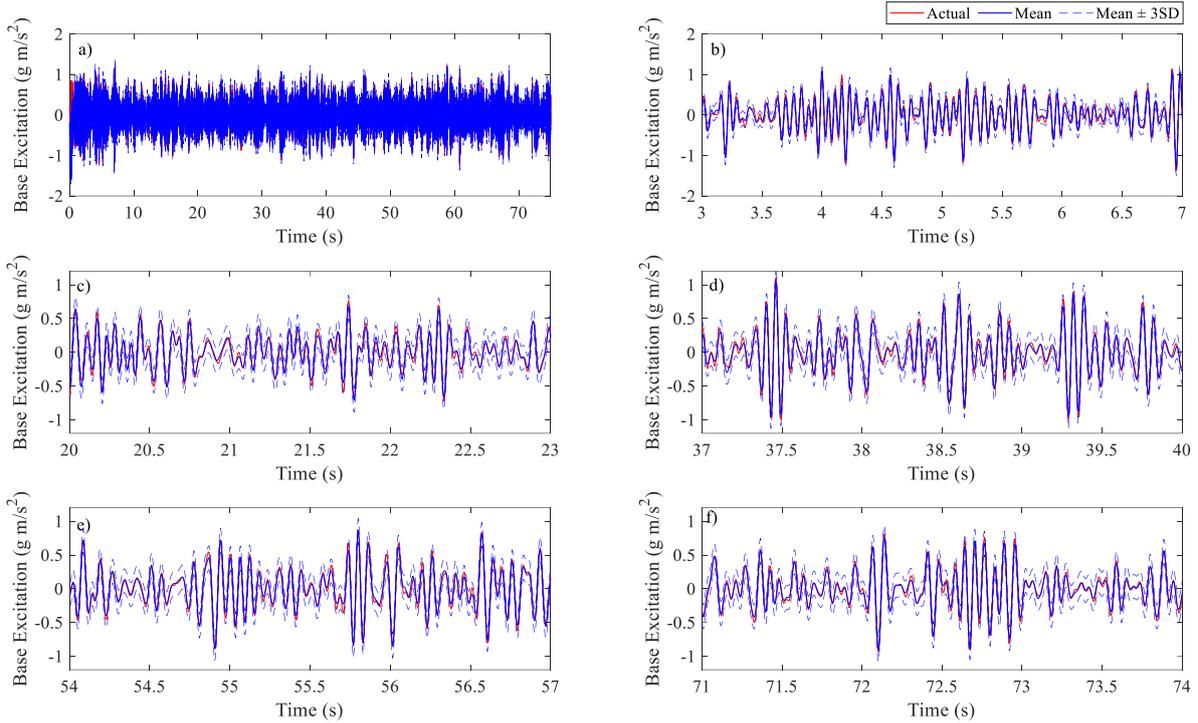

**Fig. 23.** Estimation of the base excitation using the proposed algorithm

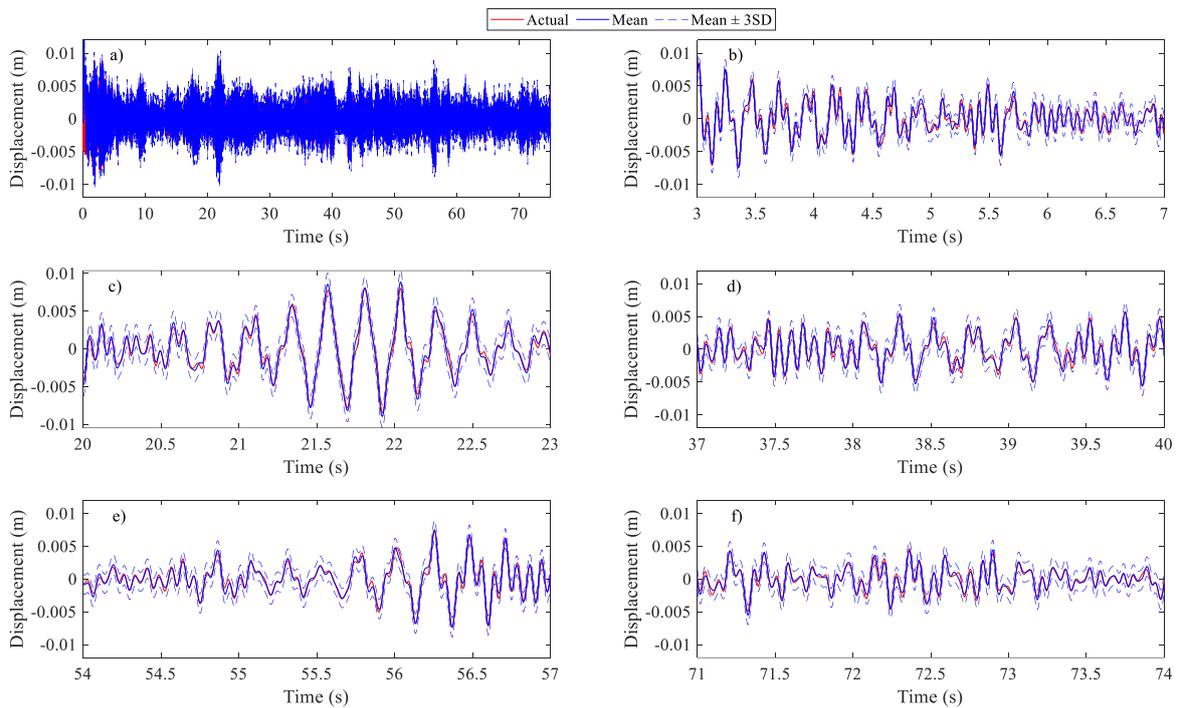

**Fig. 24.** Prediction of the displacement response of the 1st floor

Fig. 25 shows the stiffness and modal damping ratios identified using the BEM. Note that, in previous works, e.g., [57], the unknown parameters were estimated using other techniques. The agreement between the results shown in Fig. 25 and the values reported in [57] is notable. However, the dynamic behavior of the test structure is not perfectly linear, and consequently, some temporal variability prevails in the unknown parameters.

Fig. 26 shows the noise parameters over iterations of both algorithms. It is not possible to confirm the accuracy of noise parameters since no reference values are available. However, the results are deemed to be valid as long as they provide accurate estimations of the input, parameters, and states.

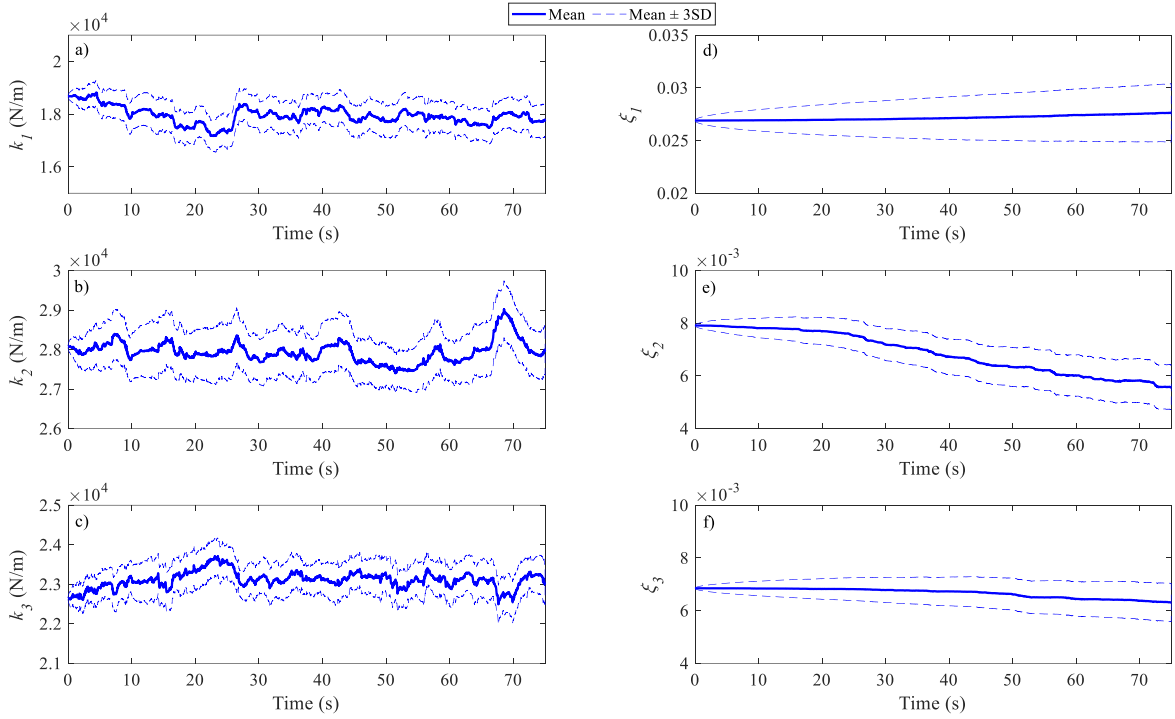

**Fig. 25.** Identification of the unknown structural parameters

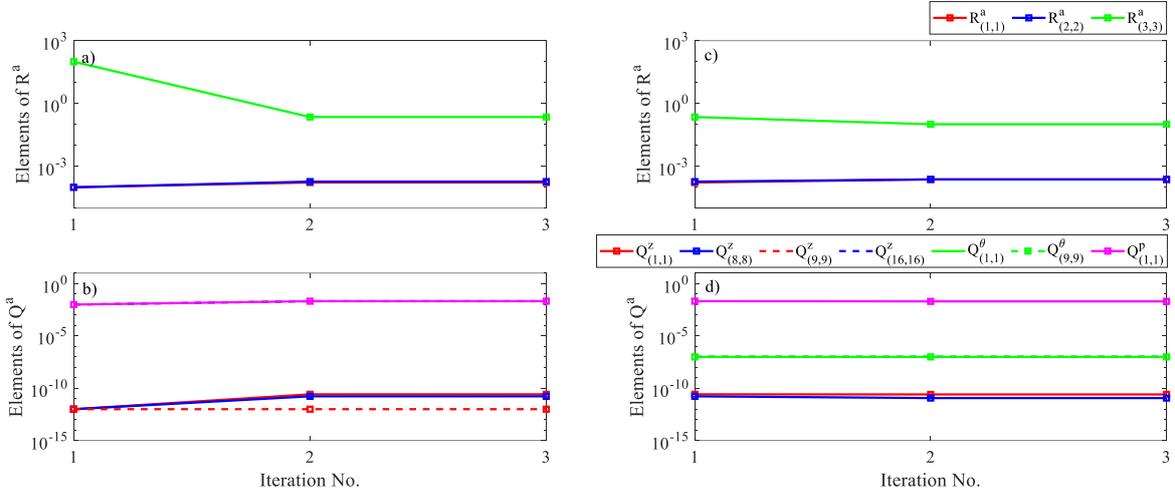

**Fig. 26.** Variations of noise parameters over iterations of (a-b) the steady-state BEM (c-d) the main BEM

## 8. Concluding Remarks and Future Work

A novel BEM methodology is proposed and validated for the identification of dynamical responses, mechanical properties, input forces, and noise characteristics while using sparse response-only measurements obtained from structures. A BEM standpoint is adopted to develop a new probabilistic formulation along with an iterative computational algorithm, which maximizes an efficient surrogate function rather than the joint posterior PDF. In limit, the optimal values obtained from the BEM tend to the MPVs of the original posterior distribution. This algorithm is efficient and easy-to-deploy as it embodies the EKF and fixed-point smoother for the estimation of all latent states while updating the noise characteristics at the end of BEM iterations. Additionally, a BEM initializer is also suggested in Appendix (B) based on the steady-state solutions of the KF and its smoother, which assists in the initial calibration of the noise parameters. Besides ensuring the stability and detectability of the system, this step significantly improves the convergence rate of the BEM. However, it could be skipped in favor of the real-time estimation of the response quantities, especially in the case sufficient information about the noise parameters is available.

Theoretical observability is also investigated through an explicit formulation, which allows for deploying a sensor network tailored for the problem at hand. However, such theoretical conditions should be viewed only as necessity for a successful structural identification, which might be inadequate in practical cases. Pseudo-observation strategies are implemented to mitigate low-frequency drifts in the estimates of states and input forces when using acceleration-only measurements. The proposed formulation and algorithms are examined using both numerical and experimental examples. Based on the results, the following conclusions can be drawn:

- The BEM can provide accurate mean estimates for input forces, responses, and parameters, when an adequate sensor configuration is used, which satisfies theoretical observability conditions. Moreover, posterior uncertainties reasonably account for discrepancies between the estimates and reference values. This desired property promotes the BEM as a powerful methodology for uncertainty quantification and propagation in system identification problems.
- The steady-state initializer is an efficient approach to acquire reasonable estimations for the process and measurement noise covariance matrices. This method can substitute the conventional methods of noise calibration. When the estimations of the noise parameters, obtained from this initializer, are fed into the BEM, the efficiency and accuracy of the BEM is further enhanced due to providing a good starting point.
- The noise updating rules do not require performing any direct optimization or imposing positive semi-definiteness conditions, as required in some methods.
- Input pseudo-observations can mitigate the low-frequency drifts in the estimations of the input and states when acceleration-only responses are measured.
- The BEM can identify abrupt changes in the stiffness of structural elements, showcasing its potential to identify structural damage and non-smooth dynamics. It is also verified that the BEM retains its accuracy in the presence of multiple unknown input forces.

In this study, the random walk model was used to describe the temporal variation of input forces. An appealing direction for further research includes investigating whether substituting this model with other stochastic processes, e.g., Gaussian process models [25,46], can improve the results. In this case, the EM algorithm can still be employed, offering consistent cross-comparison. Additionally, this paper was focused on the identification of linear dynamical systems in the presence of unknown input forces and parameters. However, the BEM is supposed to apply to nonlinear dynamical systems as well. Our efforts are underway to reformulate the BEM based on the UKF for the identification of nonlinear hysteretic systems.

**Acknowledgement**
Financial support from the Hong Kong Research Grants Council under project numbers 16212918 and 16211019 is gratefully acknowledged.

**Appendix (A).** Derivation of the observability test
Due to Eqs. (4) and (14), the joint system of process and observation models can be written in continuous-time as follows:

$$\begin{cases} \dot{\mathbf{z}}(t) = \mathbf{A}_c(\boldsymbol{\theta}(t))\mathbf{z}(t) + \mathbf{B}_c \mathbf{p}(t) \\ \mathbf{d}(t) = \mathbf{G}_c(\boldsymbol{\theta}(t))\mathbf{z}(t) + \mathbf{J}_c \mathbf{p}(t) \end{cases} \quad (A1)$$

The parameterization of the stiffness and damping matrices specified in Eqs. (2) and (3) simplifies this system of equations into:

$$\begin{cases} \dot{\mathbf{z}}(t) = \mathbf{A}_{c,0}\mathbf{z}(t) + \sum_{s=1}^{N_\theta} \theta_s(t)\mathbf{A}_{c,s}\mathbf{z}(t) + \mathbf{B}_c\mathbf{p}(t) \\ \mathbf{d}(t) = \mathbf{G}_{c,0}\mathbf{z}(t) + \sum_{s=1}^{N_\theta} \theta_s(t)\mathbf{G}_{c,s}\mathbf{z}(t) + \mathbf{J}_c\mathbf{p}(t) \end{cases} \quad (A2)$$

where $\mathbf{A}_{c,0}$ and $\mathbf{G}_{c,0}$ are the known parts of the system and observation matrices; $\mathbf{A}_{c,s}$ and $\mathbf{G}_{c,s}$ are the unknown parts of these matrices that correspond to the $s^{th}$ substructure. It can easily be verified that these matrices can be calculated from

$$\mathbf{A}_{c,s} = \begin{cases} \begin{bmatrix} \mathbf{0}_{N_d \times N_d} & \mathbf{0}_{N_d \times N_d} \\ -\mathbf{M}^{-1}\mathbf{K}_0 & -\mathbf{M}^{-1}\mathbf{C}_0 \end{bmatrix} & ; \; s = 0 \\ \begin{bmatrix} \mathbf{0}_{N_d \times N_d} & \mathbf{0}_{N_d \times N_d} \\ -\mathbf{M}^{-1}\mathbf{K}_s & \mathbf{0}_{N_d \times N_d} \end{bmatrix} & ; \; s = 1,\ldots,N_k \\ \begin{bmatrix} \mathbf{0}_{N_d \times N_d} & \mathbf{0}_{N_d \times N_d} \\ \mathbf{0}_{N_d \times N_d} & -\mathbf{M}^{-1}\mathbf{C}_s \end{bmatrix} & ; \; s = N_k+1,\ldots,N_\theta \end{cases} \quad (A4)$$

$$\mathbf{G}_{c,s} = \begin{cases} \begin{bmatrix} \mathbf{S}_d & \mathbf{0}_{n_d \times N_d} \\ -\mathbf{S}_a\mathbf{M}^{-1}\mathbf{K}_0 & -\mathbf{S}_a\mathbf{M}^{-1}\mathbf{C}_0 \end{bmatrix} & ; \; s = 0 \\ \begin{bmatrix} \mathbf{S}_d & \mathbf{0}_{n_d \times N_d} \\ -\mathbf{S}_a\mathbf{M}^{-1}\mathbf{K}_s & \mathbf{0}_{n_a \times N_d} \end{bmatrix} & ; \; s = 1,\ldots,N_k \\ \begin{bmatrix} \mathbf{S}_d & \mathbf{0}_{n_d \times N_d} \\ \mathbf{0}_{n_a \times N_d} & -\mathbf{S}_a\mathbf{M}^{-1}\mathbf{C}_s \end{bmatrix} & ; \; s = N_k+1,\ldots,N_\theta \end{cases} \quad (A5)$$

In the derivation of $\mathbf{G}_{c,s}$, only displacement and acceleration responses are considered for simplicity. Note that, when other types of sensors exist, the linear expansion specified above still holds but $\mathbf{G}_{c,s}$ needs to be reconstructed similarly. We write the first-order Taylor series expansion of Eq. (A2) around $\mathbf{z} = \mathbf{z}_0$ and $\mathbf{\theta} = \mathbf{0}$, which yields:

$$\begin{cases} \dot{\mathbf{z}}(t) = \mathbf{A}_{c,0}\mathbf{z}_0 + \mathbf{A}_{c,0}(\mathbf{z}(t) - \mathbf{z}_0) + \mathbf{B}_c\mathbf{p}(t) + \sum_{s=1}^{N_\theta} \theta_s(t)\mathbf{A}_{c,s}\mathbf{z}_0 \\ \mathbf{d}(t) = \mathbf{G}_{c,0}\mathbf{z}_0 + \mathbf{G}_{c,0}(\mathbf{z}(t) - \mathbf{z}_0) + \mathbf{J}_c\mathbf{p}(t) + \sum_{s=1}^{N_\theta} \theta_s(t)\mathbf{G}_{c,s}\mathbf{z}_0 \end{cases} \quad (A6)$$

This linearized state-space model can be rearranged into a compact form as follows:

$$\begin{cases} \dot{\mathbf{z}}(t) = \mathbf{A}_{c,0}\mathbf{z}(t) + \mathbf{B}_c\mathbf{p}(t) + \mathbf{C}_c\mathbf{\theta}(t) \\ \mathbf{d}(t) = \mathbf{G}_{c,0}\mathbf{z}(t) + \mathbf{J}_c\mathbf{p}(t) + \mathbf{H}_c\mathbf{\theta}(t) \end{cases} \quad (A7)$$

and

$$\mathbf{C}_c = \begin{bmatrix} \mathbf{A}_1\mathbf{z}_0 & \cdots & \mathbf{A}_{N_\theta}\mathbf{z}_0 \end{bmatrix} \quad (A8)$$

$$\mathbf{H}_c = \begin{bmatrix} \mathbf{G}_1\mathbf{z}_0 & \cdots & \mathbf{G}_{N_\theta}\mathbf{z}_0 \end{bmatrix} \quad (A9)$$

Based on these linearized measurement and process models, we aim to derive an explicit formulation for testing the theoretical observability. For this purpose, the procedure outlined in [51] is used. At the initial stage, the $0^{th}$-row observability is checked for the augmented state $\mathbf{z}^{(0)}(t) = [\mathbf{z}^T(t) \; \mathbf{\theta}^T(t) \; \mathbf{p}^T(t)]^T$. We further assume that $\Delta\Omega_0 = \mathbf{G}_{c,0}\mathbf{z}(t) + \mathbf{J}_c\mathbf{p}(t) + \mathbf{H}_c\mathbf{\theta}(t)$ and $\Omega_0 = \Delta\Omega_0$. Then, the derivative of $\Delta\Omega_0$ with respect to $\mathbf{z}^{(0)}(t)$ is calculated as

$$d\Omega_0 = \frac{\partial(\Delta\Omega_0)}{\partial(\mathbf{z}^{(0)}(t))^T} = \begin{bmatrix} \mathbf{G}_{c,0} & \mathbf{H}_c & \mathbf{J}_c \end{bmatrix} \quad (A10)$$

where $d\Omega_0$ governs the 0$^{\text{th}}$-row observability matrix, whose rank should be checked against $2N_d + N_\theta + N_p$. Subsequently, the 1$^{\text{st}}$-row observability requires checking the observability of $\mathbf{z}^{(1)}(t) = [(\mathbf{z}^{(0)}(t))^T \ \dot{\mathbf{p}}^T(t)]^T$ based on the Lie derivatives of $\Delta\Omega_0$ along $\mathbf{f}_{zp}^{(0)} = [\dot{\mathbf{z}}^T(t) \ \mathbf{0}^T(t) \ \dot{\mathbf{p}}^T(t)]^T$, which is calculated as

$$\Delta\Omega_1 = \frac{\partial(\Delta\Omega_0)}{\partial(\mathbf{z}^{(0)}(t))^T}\mathbf{f}_{zp}^{(0)} = \mathbf{G}_{c,0}\mathbf{A}_{c,0}\mathbf{z}(t) + (\mathbf{G}_{c,0}\mathbf{C}_c + \mathbf{H}_c)\mathbf{\theta}(t) + \mathbf{G}_{c,0}\mathbf{B}_c\mathbf{p}(t) + \mathbf{J}_c\dot{\mathbf{p}}(t) \quad (A11)$$

Then, the additional row of the observability matrix is determined from

$$d(\Delta\Omega_1) = \frac{\partial(\Delta\Omega_1)}{\partial(\mathbf{z}^{(0)}(t))^T} = \begin{bmatrix} \mathbf{G}_{c,0}\mathbf{A}_{c,0} & \mathbf{G}_{c,0}\mathbf{C}_c + \mathbf{H}_c & \mathbf{G}_{c,0}\mathbf{B}_c & \mathbf{J}_c \end{bmatrix} \quad (A12)$$

Incorporating the matrix $d(\Delta\Omega_1)$ into $d\Omega_0$ yields the 1st-row observability matrix as

$$d\Omega_1 = [d\Omega_0 \ \ \mathbf{0}] \cup d(\Delta\Omega_1) = \begin{bmatrix} \mathbf{G}_{c,0} & \mathbf{H}_c & \mathbf{J}_c & \mathbf{0} \\ \mathbf{G}_{c,0}\mathbf{A}_{c,0} & \mathbf{G}_{c,0}\mathbf{C}_c + \mathbf{H}_c & \mathbf{G}_{c,0}\mathbf{B}_c & \mathbf{J}_c \end{bmatrix} \quad (A13)$$

The 1$^{\text{st}}$-row observability condition is satisfied if the rank of $d\Omega_1$ is equal to $2N_d + N_\theta + 2N_p$. Likewise, the 2$^{\text{nd}}$-row observability should be checked for the augmented states $\mathbf{z}^{(2)}(t) = [(\mathbf{z}^{(1)}(t))^T \ \ddot{\mathbf{p}}^T(t)]^T$, which requires calculating the Lie derivatives of $\Delta\Omega_1$ along $\mathbf{f}_{zp}^{(1)} = [(\mathbf{f}_{zp}^{(0)})^T \ \ddot{\mathbf{p}}^T(t)]^T$, given by

$$\Delta\Omega_2 = \frac{\partial(\Delta\Omega_1)}{\partial(\mathbf{z}^{(1)}(t))^T}\mathbf{f}_{zp}^{(1)}$$
$$= \mathbf{G}_{c,0}\mathbf{A}_{c,0}^2\mathbf{z}(t) + (\mathbf{G}_{c,0}\mathbf{A}_{c,0}\mathbf{C}_c + \mathbf{G}_{c,0}\mathbf{C}_c + \mathbf{H}_c)\mathbf{\theta}(t) + \mathbf{G}_{c,0}\mathbf{A}_{c,0}\mathbf{B}_c\mathbf{p}(t) + \mathbf{G}_0\mathbf{B}_c\dot{\mathbf{p}}(t) + \mathbf{J}_c\ddot{\mathbf{p}}(t) \quad (A14)$$

Then, the adding part of the observability matrix is obtained as

$$d(\Delta\Omega_2) = \frac{\partial(\Delta\Omega_2)}{\partial(\mathbf{z}^{(2)}(t))^T} = \begin{bmatrix} \mathbf{G}_{c,0}\mathbf{A}_{c,0}^2 & \mathbf{G}_{c,0}\mathbf{A}_{c,0}\mathbf{C}_c + \mathbf{G}_{c,0}\mathbf{C}_c + \mathbf{H}_c & \mathbf{G}_{c,0}\mathbf{A}_{c,0}\mathbf{B}_c & \mathbf{G}_0\mathbf{B}_c & \mathbf{J}_c \end{bmatrix} \quad (A15)$$

Consequently, the 2$^{\text{nd}}$-row observability matrix is given as

$$d\Omega_2 = [d\Omega_1 \ \ \mathbf{0}] \cup d(\Delta\Omega_2)$$
$$= \begin{bmatrix} \mathbf{G}_{c,0} & \mathbf{H}_c & \mathbf{J}_c & \mathbf{0} & \mathbf{0} \\ \mathbf{G}_{c,0}\mathbf{A}_{c,0} & \mathbf{G}_{c,0}\mathbf{C}_c + \mathbf{H}_c & \mathbf{G}_{c,0}\mathbf{B}_c & \mathbf{J}_c & \mathbf{0} \\ \mathbf{G}_{c,0}\mathbf{A}_{c,0}^2 & \mathbf{G}_{c,0}\mathbf{A}_{c,0}\mathbf{C}_c + \mathbf{G}_{c,0}\mathbf{C}_c + \mathbf{H}_c & \mathbf{G}_{c,0}\mathbf{A}_{c,0}\mathbf{B}_c & \mathbf{G}_{c,0}\mathbf{B}_c & \mathbf{J}_c \end{bmatrix} \quad (A16)$$

Continuing this process allows deriving a general formulation for the $k^{\text{th}}$-row observability matrix, which leads to

$$d\Omega_k = \begin{bmatrix} \mathbf{G}_{c,0} & \mathbf{H}_c & \mathbf{J}_c & \mathbf{0} & \mathbf{0} & \mathbf{0} & \mathbf{0} \\ \mathbf{G}_{c,0}\mathbf{A}_{c,0} & \mathbf{G}_{c,0}\mathbf{C}_c + \mathbf{H}_c & \mathbf{G}_{c,0}\mathbf{B}_c & \mathbf{J}_c & \mathbf{0} & \mathbf{0} & \mathbf{0} \\ \mathbf{G}_{c,0}\mathbf{A}_{c,0}^2 & \mathbf{G}_{c,0}\mathbf{A}_{c,0}\mathbf{C}_c + \mathbf{G}_{c,0}\mathbf{C}_c + \mathbf{H}_c & \mathbf{G}_{c,0}\mathbf{A}_{c,0}\mathbf{B}_c & \mathbf{G}_{c,0}\mathbf{B}_c & \mathbf{J}_c & \mathbf{0} & \mathbf{0} \\ \vdots & \vdots & \vdots & \ddots & \ddots & \ddots & \mathbf{0} \\ \mathbf{G}_{c,0}\mathbf{A}_{c,0}^k & \sum_{j=0}^{k-1}\mathbf{G}_{c,0}\mathbf{A}_{c,0}^j\mathbf{C}_c + \mathbf{H}_c & \mathbf{G}_{c,0}\mathbf{A}_{c,0}^{k-1}\mathbf{B}_c & \cdots & \mathbf{G}_{c,0}\mathbf{A}_{c,0}\mathbf{B}_c & \mathbf{G}_{c,0}\mathbf{B}_c & \mathbf{J}_c \end{bmatrix} \quad (A17)$$

If the rank of this matrix is equal to $2N_d + N_\theta + (k+1)N_p$, the $k^{\text{th}}$-row observability is satisfied.

**Appendix (B). Steady-state EM Initializer**

In this appendix, an initialization strategy is introduced for Algorithm 1, which uses asymptotic properties of the BEM methodology when the structural parameters are fixed at some estimation. This strategy allows acquiring a sense of the measurement and process noise characteristics at least in terms of their order of magnitude through the steady-state solutions of the Bayesian estimators.

For LTI systems with no unknown parameters, the posterior covariance matrix of the state vector stabilizes as the number of data points increases. This condition can be assessed through the steady-state solution of the estimators, which can be attained when the system and observation matrices are known. Nevertheless, the derivation of the steady-state solution is a non-trivial process when the system is known partially. Since the initial estimation of the noise parameters are required only when starting the main EM algorithm, it is reasonable to set the structural parameters at a rational value, e.g., nominal values ($\boldsymbol{\theta}_k = \boldsymbol{\theta}_0, \forall k \in \{0,1,2,...,n\}$). In this case, the structural parameters will be removed from the augmented state vector, giving the following process and observation models:

$$\boldsymbol{\zeta}_k = \mathbf{A}^\zeta \boldsymbol{\zeta}_{k-1} + \mathbf{v}_k^\zeta \tag{B1}$$

$$\mathbf{d}_k^a = \mathbf{G}^\zeta \boldsymbol{\zeta}_k + \mathbf{w}_k^a \tag{B2}$$

where $\boldsymbol{\zeta}_k = [(\mathbf{z}_k)^T \ (\mathbf{p}_k)^T]^T$ is the reduced state vector comprising $N_\zeta = 2N_d + N_p$ unknown elements; $\mathbf{v}_k^\zeta = [(\mathbf{v}_k^z)^T \ (\mathbf{v}_k^p)^T]^T$ is the reduced process noise, considered to be zero-mean GWN with $\mathbf{Q}^\zeta = \text{block-diag}[\mathbf{Q}^z, \mathbf{Q}^p]$ covariance matrix; $\mathbf{A}^\zeta$ and $\mathbf{G}^\zeta$ are time-invariant system and observation matrices, calculated as

$$\mathbf{A}^\zeta = \begin{bmatrix} \mathbf{A}(\boldsymbol{\theta}_0) & \mathbf{B}(\boldsymbol{\theta}_0) \\ \mathbf{0}_{N_p \times 2N_d} & \mathbf{I}_{N_p \times N_p} \end{bmatrix} \quad ; \quad \mathbf{G}^\zeta = \begin{bmatrix} \mathbf{G}_c(\boldsymbol{\theta}_0) & \mathbf{J}_c \end{bmatrix} \tag{B3}$$

For this probabilistic setting, the joint posterior PDF can be derived similar to Eq. (32), giving:

$$p(\{\boldsymbol{\zeta}_k\}_{k=0}^n, \boldsymbol{\mu}_{\zeta_0}, \mathbf{P}_{\zeta_0}, \mathbf{Q}^\zeta, \mathbf{R}^a \mid D_n) \propto N(\boldsymbol{\zeta}_0 \mid \boldsymbol{\mu}_{\zeta_0}, \mathbf{P}_{\zeta_0}) \prod_{k=1}^n N(\mathbf{d}_k^a \mid \mathbf{G}^\zeta \boldsymbol{\zeta}_k, \mathbf{R}^a) N(\boldsymbol{\zeta}_k \mid \mathbf{A}^\zeta \boldsymbol{\zeta}_{k-1}, \mathbf{Q}^\zeta) \tag{B4}$$

where $\boldsymbol{\mu}_{\zeta_0} \in \mathbb{R}^{N_\zeta}$ and $\mathbf{P}_{\zeta_0} \in \mathbb{R}^{N_\zeta \times N_\zeta}$ denote the mean and covariance of the reduced state vector, which should be identified as well. Since this posterior PDF has the same structure as Eq. (32), the proposed BEM methodology applies albeit with slight adjustments. Specifically, the matrices $\mathbf{F}_{k-1|k-1}^\xi$ and $\mathbf{H}_{k|k-1}^\xi$ should be replaced by time-invariant quantities $\mathbf{A}^\zeta$ and $\mathbf{G}^\zeta$, respectively. By doing so, the posterior distribution can be rewritten as

$$p(Z_n, \Lambda \mid D_n) \propto N(\boldsymbol{\zeta}_0 \mid \boldsymbol{\mu}_{\zeta_0}, \mathbf{P}_{\zeta_0}) \prod_{k=1}^n N(\mathbf{d}_k^a \mid \mathbf{G}^\zeta \boldsymbol{\zeta}_k, \mathbf{R}^a) N(\boldsymbol{\zeta}_k \mid \mathbf{A}^\zeta \boldsymbol{\zeta}_{k-1}, \mathbf{Q}^\zeta) \tag{B5}$$

where $Z_n = \{\boldsymbol{\zeta}_k\}_{k=0}^n$ is a set comprising all latent states; $\Lambda = \{\boldsymbol{\mu}_{\zeta_0}, \mathbf{P}_{\zeta_0}, \mathbf{Q}^\zeta, \mathbf{R}^a\}$ is a set comprising the unknown hyper-parameters of the probabilistic model. For this probability distribution, the $j$[th] iteration of the EM algorithm applies as

E-Step: Calculate $L(\Lambda \mid \hat{\Lambda}^{(j-1)}) = \mathbb{E}_{Z_n}[\ln p(Z_n, \Lambda \mid D_n)]$ (B6)

M-Step: Maximize $L(\Lambda \mid \hat{\Lambda}^{(j-1)})$ with respect to $\Lambda$ (B7)

where the surrogate function reads as

$$L(\Lambda \mid \hat{\Lambda}^{(j-1)}) = -\frac{n}{2}\ln|\mathbf{R}^a| - \frac{n}{2}\ln|\mathbf{Q}^\zeta| - \frac{1}{2}\ln|\mathbf{P}_{\zeta_0}| + \frac{1}{2}tr\left((\mathbf{P}_{\zeta_0})^{-1}\mathbb{E}_{\zeta_0}[(\boldsymbol{\zeta}_0 - \boldsymbol{\mu}_{\zeta_0})(\boldsymbol{\zeta}_0 - \boldsymbol{\mu}_{\zeta_0})^T]\right)$$
$$+\frac{1}{2}tr\left((\mathbf{R}^a)^{-1}\sum_{k=1}^n \mathbb{E}_{\zeta_k}\left[(\mathbf{d}_k^a - \mathbf{G}^\zeta \boldsymbol{\zeta}_k)(\mathbf{d}_k^a - \mathbf{G}^\zeta \boldsymbol{\zeta}_k)^T\right] + (\mathbf{Q}^\zeta)^{-1}\sum_{k=1}^n \mathbb{E}_{\zeta_k,\zeta_{k-1}}\left[(\boldsymbol{\zeta}_k - \mathbf{A}^\zeta \boldsymbol{\zeta}_{k-1})(\boldsymbol{\zeta}_k - \mathbf{A}^\zeta \boldsymbol{\zeta}_{k-1})^T\right]\right)$$
(B8)

where $\hat{\Lambda}^{(j-1)}$ is the optimal values of the hyper-parameters obtained at the ($j$-1)[th] iteration. Then, the Bayesian estimator obtained in Eqs. (39-43) can be reformulated as

$$\boldsymbol{\zeta}_{k|k-1} = \mathbf{A}^{\zeta}\boldsymbol{\zeta}_{k-1|k-1}$$

$$\mathbf{P}^{\zeta}_{k|k-1} = \mathbf{A}^{\zeta}\mathbf{P}^{\zeta}_{k-1|k-1}(\mathbf{A}^{\zeta})^{T} + \mathbf{Q}^{\zeta}$$

$$\mathbf{G}^{\zeta}_{k} = \mathbf{P}^{\zeta}_{k|k-1}(\mathbf{G}^{\zeta})^{T}(\mathbf{R}^{a} + \mathbf{G}^{\zeta}\mathbf{P}^{\zeta}_{k|k-1}(\mathbf{G}^{\zeta})^{T})^{-1} \quad (B9)$$

$$\boldsymbol{\zeta}_{k|k} = \boldsymbol{\zeta}_{k|k-1} + \mathbf{G}^{\zeta}_{k}(\mathbf{d}^{a}_{k} - \mathbf{G}^{\zeta}\boldsymbol{\zeta}_{k|k-1})$$

$$\mathbf{P}^{\zeta}_{k|k} = \mathbf{P}^{\zeta}_{k|k-1} - \mathbf{G}^{\zeta}_{k}\mathbf{G}^{\zeta}\mathbf{P}^{\zeta}_{k|k-1}$$

where $\boldsymbol{\zeta}_{k|k-1} \in \mathbb{R}^{N_{\varsigma}}$ and $\mathbf{P}^{\zeta}_{k|k-1} \in \mathbb{R}^{N_{\varsigma} \times N_{\varsigma}}$ are the predictive mean and covariance of $\boldsymbol{\zeta}_{k}$, respectively; $\boldsymbol{\zeta}_{k|k} \in \mathbb{R}^{N_{\varsigma}}$ and $\mathbf{P}^{\zeta}_{k|k} \in \mathbb{R}^{N_{\varsigma} \times N_{\varsigma}}$ are the updated mean and covariance of $\boldsymbol{\zeta}_{k}$, respectively; $\mathbf{G}^{\zeta}_{k}$ is the estimator gain matrix. In the same vein, the fixed-point smoother can be rewritten as

$$\mathbf{L}^{\zeta}_{k} = \mathbf{P}^{\zeta}_{k|k}(\mathbf{A}^{\zeta})^{T}(\mathbf{P}^{\zeta}_{k+1|k})^{-1}$$

$$\boldsymbol{\zeta}_{k|k+1} = \boldsymbol{\zeta}_{k|k} + \mathbf{L}^{\zeta}_{k}(\boldsymbol{\zeta}_{k+1|k+1} - \boldsymbol{\zeta}_{k+1|k}) \quad (B10)$$

$$\mathbf{P}^{\zeta}_{k|k+1} = \mathbf{P}^{\zeta}_{k|k} + \mathbf{L}^{\zeta}_{k}(\mathbf{P}^{\zeta}_{k+1|k+1} - \mathbf{P}^{\zeta}_{k+1|k})(\mathbf{L}^{\zeta}_{k})^{T}$$

where $\boldsymbol{\zeta}_{k|k+1} \in \mathbb{R}^{N_{\varsigma}}$ and $\mathbf{P}^{\zeta}_{k|k+1} \in \mathbb{R}^{N_{\varsigma} \times N_{\varsigma}}$ are the smoothed mean and covariance of the reduced state, respectively; $\mathbf{L}^{\zeta}_{k} \in \mathbb{R}^{N_{\varsigma} \times N_{\varsigma}}$ is the smoother gain matrix. This formulation provides recursive estimates of the dynamical states and input forces when the structural parameters are set to predefined fixed values. Moreover, it allows proposing an EM initializer based on the stationary solutions of the Bayesian filter and smoother, as obtained next.

To derive the stationary solutions, the predictive state covariance matrix is assumed to stabilize for a large number of data points, i.e., $\lim_{k \to \infty} \mathbf{P}^{\zeta}_{k+1|k} = \lim_{k \to \infty} \mathbf{P}^{\zeta}_{k|k-1} = \mathbf{P}^{\zeta-}_{\infty}$. By applying this condition to Eq. (B5), we will arrive at the following algebraic Riccati equation:

$$\mathbf{P}^{\zeta-}_{\infty} = \mathbf{A}^{\zeta}\mathbf{P}^{\zeta-}_{\infty}(\mathbf{A}^{\zeta})^{T} - \mathbf{A}^{\zeta}\mathbf{P}^{\zeta-}_{\infty}(\mathbf{G}^{\zeta})^{T}(\mathbf{R}^{a} + \mathbf{G}^{\zeta}\mathbf{P}^{\zeta-}_{\infty}(\mathbf{G}^{\zeta})^{T})^{-1}\mathbf{G}^{\zeta}\mathbf{P}^{\zeta-}_{\infty}(\mathbf{A}^{\zeta})^{T} + \mathbf{Q}^{\zeta} \quad (B11)$$

where $\mathbf{P}^{\zeta-}_{\infty} \in \mathbb{R}^{N_{\varsigma} \times N_{\varsigma}}$ is the stationary predictive covariance matrix of the reduced state, which can be calculated using "*idare*" in MATLAB [58]. Once $\mathbf{P}^{\zeta-}_{\infty}$ is obtained, one can write:

$$\mathbf{K}^{\zeta}_{\infty} = \mathbf{P}^{\zeta-}_{\infty}(\mathbf{G}^{\zeta})^{T}(\mathbf{R}^{a} + \mathbf{G}^{\zeta}\mathbf{P}^{\zeta-}_{\infty}(\mathbf{G}^{\zeta})^{T})^{-1} \quad (B12)$$

$$\mathbf{P}^{\zeta}_{\infty} = \mathbf{P}^{\zeta-}_{\infty} - \mathbf{K}^{\zeta}_{\infty}\mathbf{G}^{\zeta}\mathbf{P}^{\zeta-}_{\infty} \quad (B13)$$

where $\mathbf{K}^{\zeta}_{\infty} \in \mathbb{R}^{N_{\varsigma} \times (N_m + N_{pd})}$ and $\mathbf{P}^{\zeta}_{\infty} \in \mathbb{R}^{N_{\varsigma} \times N_{\varsigma}}$ are respectively the stationary gain and covariance matrices. Since the gain matrix is constant, the reduced state vector can recursively be estimated as

$$\boldsymbol{\zeta}^{\infty}_{k|k} = \mathbf{A}^{\zeta}\boldsymbol{\zeta}^{\infty}_{k-1|k-1} + \mathbf{K}^{\zeta}_{\infty}(\mathbf{d}^{a}_{k} - \mathbf{G}^{\zeta}\mathbf{A}^{\zeta}\boldsymbol{\zeta}^{\infty}_{k-1|k-1}) \quad (B14)$$

where $\boldsymbol{\zeta}^{\infty}_{k|k} \in \mathbb{R}^{N_{\varsigma}}$ is the steady-state estimation of $\boldsymbol{\zeta}_{k}$ conditional on $D_{k}$. It should be noted that a sufficient condition for the existence of at least one positive semi-definite solution for the discrete-time Riccati equation is linked to the system detectability [58]. Thus, an adequate sensor configuration which makes these steady-state estimations attainable is an underlying assumption, required to be considered.

A similar formulation can be obtained for the Bayesian smoother, characterized in Eq. (B10). As $\mathbf{P}^{\zeta-}_{\infty}$ is given, the smoothing gain matrix can be computed as

$$\mathbf{L}^{\zeta}_{\infty} = \mathbf{P}^{\zeta}_{\infty}(\mathbf{A}^{\zeta})^{T}(\mathbf{P}^{\zeta-}_{\infty})^{-1} \quad (B15)$$

For a large number of data points, the fixed-point smoothing covariance matrix also stabilizes in the sense that $\lim_{k \to \infty} \mathbf{P}^{\zeta}_{k|k+1} = \lim_{k \to \infty} \mathbf{P}^{\zeta}_{k-1|k} = \mathbf{P}^{\zeta s}_{\infty}$. Substituting this condition into Eq. (B10) yields the following Lyapunov equation:

$$\mathbf{L}^{\zeta}_{\infty}\mathbf{P}^{\zeta s}_{\infty}(\mathbf{L}^{\zeta}_{\infty})^{T} - \mathbf{P}^{\zeta s}_{\infty} + (\mathbf{P}^{\zeta}_{\infty} - \mathbf{L}^{s}_{\infty}\mathbf{P}^{\zeta-}_{\infty}(\mathbf{L}^{\zeta}_{\infty})^{T}) = \mathbf{0} \quad (B16)$$

where $\mathbf{P}_\infty^{\zeta s} \in \mathbb{R}^{N_\varsigma \times N_\varsigma}$ is the state covariance matrix smoothed through the steady-state solutions. The solution of this equation can easily be obtained using "*dylap*" in MATLAB [58]. Once $\mathbf{P}_\infty^{\zeta s}$ is obtained, the state can be smoothed as

$$\zeta_{k|k+1}^\infty = \zeta_{k|k}^\infty + \mathbf{L}_\infty^\zeta (\zeta_{k+1|k+1}^\infty - \mathbf{A}^\zeta \zeta_{k|k}^\infty) \tag{B17}$$

where $\zeta_{k|k+1}^\infty \in \mathbb{R}^{N_\varsigma}$ is the smoothed state vector. Based on these steady-state estimations, the initial conditions, as well as the noise covariance matrices can be calculated in a fashion similar to the process followed in Section 4.2, giving:

$$\hat{\boldsymbol{\mu}}_{\zeta_0}^\infty = \zeta_{0|1}^\infty \tag{B18}$$

$$\hat{\mathbf{P}}_{\zeta_0}^\infty = \mathbf{P}_\infty^{\zeta s} \tag{B19}$$

$$\hat{\mathbf{R}}_\infty^a = \frac{1}{n} \sum_{k=1}^{n} \left[ (\mathbf{d}_k^a - \mathbf{G}^\zeta \zeta_{k|k+1}^\infty)(\mathbf{d}_k^a - \mathbf{G}^\zeta \zeta_{k|k+1}^\infty)^T \right] + \mathbf{G}^\zeta \mathbf{P}_\infty^{\zeta s} (\mathbf{G}^\zeta)^T \tag{B20}$$

$$\hat{\mathbf{Q}}_\infty^\zeta = \frac{1}{n} \sum_{k=1}^{n} \left[ (\zeta_{k|k+1}^\infty - \mathbf{A}^\zeta \zeta_{k-1|k}^\infty)(\zeta_{k|k+1}^\infty - \mathbf{A}^\zeta \zeta_{k-1|k}^\infty)^T \right] + \mathbf{P}_\infty^{\zeta s} + \mathbf{A}^\zeta \mathbf{P}_\infty^{\zeta s} (\mathbf{A}^\zeta)^T - \mathbf{A}^\zeta \mathbf{L}_\infty^\zeta \mathbf{P}_\infty^{\zeta s} - \mathbf{P}_\infty^{\zeta s} (\mathbf{L}_\infty^\zeta)^T (\mathbf{A}^\zeta)^T \tag{B21}$$

where $\hat{\boldsymbol{\mu}}_{\zeta_0}^\infty \in \mathbb{R}^{N_\varsigma}$ and $\hat{\mathbf{P}}_{\zeta_0}^\infty \in \mathbb{R}^{N_\varsigma \times N_\varsigma}$ are the steady-state estimation of the mean and covariance of $\zeta_0 \in \mathbb{R}^{N_\varsigma}$, respectively; $\hat{\mathbf{R}}_\infty^a$ and $\hat{\mathbf{Q}}_\infty^\zeta$ are stationary estimations of the measurement and process noise covariance matrices, respectively. These estimations can be used when starting the Algorithm 1. At the worst case scenario, when the nominal values of the structural parameters are not so accurate, these estimations are still useful for estimating the order of magnitude of the noise parameters. Moreover, this steady-state approach requires a shorter run-time compared to the main EM algorithm since the recursive calculation of the structural parameters, gain, and covariance matrices is not required. Therefore, when it is used before running the main algorithm, it can help reduce the number of iterations and the total run time of the main algorithm.

Algorithm 2 puts together the steady-state EM initializer. At first, the mean and covariance of the initial state and input should be provided, which can be considered zero. Then, an estimation of the structural parameters is needed for starting the optimization. The same should be done for both $\mathbf{Q}^z$ and $\mathbf{Q}^p$, which can be addressed based on users' judgement or any other conventional approach. The while-loop constitutes the main part of Algorithm 2, which recursively provides stationary estimates of the state and input for predefined structural parameters. At the end of the iteration, the noise covariance matrices can be computed using the M-Step's formulation. This procedure should be repeated until the same convergence criteria as Algorithm 1 are satisfied. Ultimately, updated values of $\mathbf{Q}^z$ and $\mathbf{Q}^p$ will be obtained, which can be used when running Algorithm 1.

**Algorithm 2.**
Steady-state BEM initializer for tuning noise covariance matrices

1: Set the initial state mean and covariance ($\boldsymbol{\mu}_{\zeta_0}$ and $\mathbf{P}_{\zeta_0}$).

2: Set a rough estimate for the structural parameters ($\boldsymbol{\theta}_0$).

3: Calculate the system and observation matrices ($\mathbf{A}^\zeta$ and $\mathbf{G}^\zeta$) based on Eq. (B3).

4: Set the noise covariance matrices $\hat{\mathbf{Q}}_\infty^\zeta$ and $\hat{\mathbf{R}}_\infty^a$ (Any rules of thumb may be applied).

5: Set the convergence tolerance ($TOL$), e.g., $10^{-4}$; the maximum number of iterations ($ITRMAX$), e.g., 200.

6: Set the iteration number ($j$), the convergence metric ($CON$), and the surrogate function ($\hat{L}_0$) to 1.

7: **While** ($CON < TOL$) **or** ($j < ITRMAX$) {

8:     **E-Step:** Calculate the steady-state estimation of the condensed state vector.

9:     Solve the following Riccati equation and calculate the stationary predictive covariance matrix ($\mathbf{P}_\infty^{\zeta-}$)

10:     $\mathbf{P}_\infty^{\zeta-} = \mathbf{A}^\zeta \mathbf{P}_\infty^{\zeta-} (\mathbf{A}^\zeta)^T - \mathbf{A}^\zeta \mathbf{P}_\infty^{\zeta-} (\mathbf{G}^\zeta)^T (\hat{\mathbf{R}}_\infty^a + \mathbf{G}^\zeta \mathbf{P}_\infty^{\zeta-} (\mathbf{G}^\zeta)^T)^{-1} \mathbf{G}^\zeta \mathbf{P}_\infty^{\zeta-} (\mathbf{A}^\zeta)^T + \hat{\mathbf{Q}}_\infty^\zeta$

11:     Calculate the stationary filtering gain: $\mathbf{K}_\infty^\zeta = \mathbf{P}_\infty^{\zeta-} (\mathbf{G}^\zeta)^T (\hat{\mathbf{R}}_\infty^a + \mathbf{G}^\zeta \mathbf{P}_\infty^{\zeta-} (\mathbf{G}^\zeta)^T)^{-1}$

12:     Calculate the stationary covariance matrix of state: $\mathbf{P}_\infty^\zeta = \mathbf{P}_\infty^{\zeta-} - \mathbf{K}_\infty^\zeta \mathbf{G}^\zeta \mathbf{P}_\infty^{\zeta-}$

13:     Calculate the stationary smoothing gain: $\mathbf{L}_\infty^\zeta = \mathbf{P}_\infty^\zeta (\mathbf{A}^\zeta)^T (\mathbf{P}_\infty^{\zeta-})^{-1}$

14:     Solve the Lyapunov equation for the stationary smoothed-covariance matrix of state ($\mathbf{P}_\infty^s$)

15:     $\mathbf{L}_\infty^\zeta \mathbf{P}_\infty^{\zeta s} (\mathbf{L}_\infty^\zeta)^T - \mathbf{P}_\infty^{\zeta s} + (\mathbf{P}_\infty^\zeta - \mathbf{L}_\infty^s \mathbf{P}_\infty^{\zeta-} (\mathbf{L}_\infty^\zeta)^T) = \mathbf{0}$

16:     **For** $k = 1:n$ {Stationary fixed-point smoothing:

17:         $\boldsymbol{\zeta}_{k|k}^\infty = \mathbf{A}^\zeta \boldsymbol{\zeta}_{k-1|k-1}^\infty + \mathbf{K}_\infty^\zeta (\mathbf{d}_k^a - \mathbf{G}^\zeta \mathbf{A}^\zeta \boldsymbol{\zeta}_{k-1|k-1}^\infty)$

        $\boldsymbol{\zeta}_{k-1|k}^\infty = \boldsymbol{\zeta}_{k-1|k-1}^\infty + \mathbf{L}_\infty^\zeta (\boldsymbol{\zeta}_{k|k}^\infty - \mathbf{A}^\zeta \boldsymbol{\zeta}_{k-1|k-1}^\infty)$

18:     } **End For**

19: M-Step: update the noise covariance matrices and the initial conditions parameters using Eqs. (B18-B21)

20: $\hat{\boldsymbol{\mu}}_{\zeta_0}^\infty = \boldsymbol{\zeta}_{0|1}^\infty$

21: $\hat{\mathbf{P}}_{\zeta_0}^\infty = \mathbf{P}_\infty^{\zeta s}$

22: $\hat{\mathbf{R}}_\infty^a = \dfrac{1}{n} \sum_{k=1}^n \left[ (\mathbf{d}_k^a - \mathbf{G}^\zeta \boldsymbol{\zeta}_{k|k+1}^\infty)(\mathbf{d}_k^a - \mathbf{G}^\zeta \boldsymbol{\zeta}_{k|k+1}^\infty)^T \right] + \mathbf{G}^\zeta \mathbf{P}_\infty^{\zeta s} (\mathbf{G}^\zeta)^T$

23: $\hat{\mathbf{Q}}_\infty^\zeta = \dfrac{1}{n} \sum_{k=1}^n \left[ (\boldsymbol{\zeta}_{k|k+1}^\infty - \mathbf{A}^\zeta \boldsymbol{\zeta}_{k-1|k}^\infty)(\boldsymbol{\zeta}_{k|k+1}^\infty - \mathbf{A}^\zeta \boldsymbol{\zeta}_{k-1|k}^\infty)^T \right] + \mathbf{P}_\infty^{\zeta s} + \mathbf{A}^\zeta \mathbf{P}_\infty^{\zeta s} (\mathbf{A}^\zeta)^T - \mathbf{A}^\zeta \mathbf{L}_\infty^\zeta \mathbf{P}_\infty^{\zeta s} - \mathbf{P}_\infty^{\zeta s} (\mathbf{L}_\infty^\zeta)^T (\mathbf{A}^\zeta)^T$

24: Calculate $\hat{L}_1 = L(\Lambda | \hat{\Lambda}^{(j-1)})$ based on Eq. (B8)

25: Calculate $CON = (\hat{L}_1 - \hat{L}_0) / \hat{L}_0$

26: Set $\hat{L}_0 = \hat{L}_1$ and $j = j+1$.

27: } **End While**